\pdfoutput=1
\documentclass[fleqn]{elsart}
\usepackage{graphicx,color}
\usepackage{amsmath,amssymb}
\usepackage{slashbox}

\def\anp#1#2#3{Annals Phys. #1 (#3) #2}

\def\ibid#1#2#3{{\it ibid.} #1 (#3) #2}
\def\ijma#1#2#3{Intl. Jour. Mod. Phys. A #1 (#3) #2}

\def\jhep#1#2#3{J. High Energy Phys. #2 (#3) #1}

\def\jpg#1#2#3{Jour. Phys. G #1 (#3) #2}

\def\npa#1#2#3{Nucl. Phys. A #1 (#3) #2}
\def\npb#1#2#3{Nucl. Phys. B #1 (#3) #2}

\def\plb#1#2#3{Phys. Lett. B #1 (#3) #2}

\def\prd#1#2#3{Phys. Rev. D #1 (#3) #2}
\def\prl#1#2#3{Phys. Rev. Lett. #1 (#3) #2}
\def\phr#1#2#3{Phys. Rep. #1 (#3) #2}

\def\ptp#1#2#3{Prog. Theor. Phys. #1 (#3) #2}

\newcommand{\be}{\begin{equation}}
\newcommand{\ee}{\end{equation}}
\newcommand{\bea}{\begin{eqnarray}}
\newcommand{\eea}{\end{eqnarray}}
\newcommand{\Slash}[1]{{\ooalign{\hfil/\hfil\crcr$#1$}}}
\newcommand{\tr}{{\rm tr}}
\newcommand{\Nc}{N_{\rm c}}
\newcommand{\Nf}{N_{\rm f}}

\newcommand{\lqcd}{\Lambda_{\rm QCD}}

\newcommand{\la}{\langle}
\newcommand{\ra}{\rangle}
\newcommand{\eff}{ {\rm eff} }
\newcommand{\qcd}{QCD$_2$\ }

\newcommand{\rmd}{\mathrm{d}}
\newcommand{\rmi}{\mathrm{i}}
\newcommand{\rme}{\mathrm{e}}
\newcommand{\rmA}{\mathrm{A}}
\newcommand{\rmB}{\mathrm{B}}

\begin{document}
\begin{flushright}
{\scriptsize RBRC 881}
\end{flushright}
\begin{frontmatter}
\title{A (1+1) dimensional example of Quarkyonic matter}
\author{Toru Kojo}
\address{RIKEN BNL Research Center, Brookhaven National
  Laboratory,\\ Upton, NY-11973, USA}
\begin{abstract}
We analyze the (1+1) dimensional QCD (QCD$_2$) at finite
density to consider a number of qualitative issues:
confinement in dense quark matter,
the chiral symmetry breaking near the Fermi surface,
the relation between chiral spirals
and quark number density,
and a possibility of the spontaneous 
flavor symmetry breaking.
We argue that while
the free energy is dominated by perturbative quarks,
confined excitations at zero density can persist up to high density.
So quark matter in QCD$_2$ is an example of 
Quarkyonic matter.
The non-Abelian bosonization
and associated charge-flavor-color
separation are mainly used in order to clarify basic structures
of QCD$_2$ at finite density.
\end{abstract}
\end{frontmatter}
\section{Introduction}
\label{intro}
The confinement is a distinct phenomenon in
Quantum Chromodynamics (QCD).
The confinement reveals its property in
the physical spectra saturated by
the color singlet states.
Many features of spectra are qualitatively explained
by the picture of the squeezed colored 
flux connecting colored objects.

Recently, the role of confinement in 
cold, dense quark matter has been addressed
by McLerran and Pisarski \cite{McLerran:2007qj}.
They suggested that
excitations can remain confined
even after quarks are released from baryons
to form quark matter.
Such quark matter is named 
Quarkyonic matter 
\cite{McLerran:2007qj,McLerran:2008ua,Kojo:2009ha,Kojo:2010fe},
which is distinguished from the conventional
quark matter with deconfined excitations.
The understanding of excitation modes
is important to consider the phase structure,
transport phenomena, etc.

Before suggestions in Ref.\cite{McLerran:2007qj}, 
confining effects in quark matter
have not been taken into account seriously. 
Presumably one of the reasons would be that
quarks seem to excite individually
after being released from baryons.
In addition, asymptotic freedom apparently
implies that
confining forces become irrelevant
when the typical distance among quarks
is very small.

In this paper, we are going
to offer one counter example to such reasonings.
While we expect that conventional reasonings work
for most of regions in the quark Fermi sea,
they do not reflect physics near the Fermi surface.
To illustrate the points, 
we analyze QCD in (1+1) dimensions (QCD$_2$) 
as a theory with confining forces and asymptotic freedom
\cite{'tHooft:1974hx,Callan:1975ps,Colemanbook,Affleck:1985wa,Zhitnitsky:1985um,Salcedo:1990rw}.
We will conclude that cold, dense quark matter in QCD$_2$
is an example of Quarkyonic matter.

Our main statement is that 
the mechanism to suppress the confining gluons 
at finite density
is neither asymptotic freedom nor
the percolation of confined bags, 
but the medium induced color screening.
Since the color charge of the system is zero in average,
the strength of the screening is determined by
the virtual processes.
The main actors are the virtual colored fluctuations
with low energy:
quarks near the Fermi surface.
The contributions from virtual quark fluctuations are enhanced
as the area of the quark Fermi surface grows.
In spatial dimensions, $d$, larger than one,
the phase space for quark fluctuations 
increases as $\sim \mu^{d-1} \lqcd$,
so the strength of the screening grows as 
$\sim (\mu/\lqcd)^{d-1}$, compared to the vacuum case
($\mu,\lqcd$ are quark chemical potential
and nonperturbative scale, respectively).
So deconfinement should take place when
quark fluctuations become comparable to those of gluons,
\begin{equation}
\Nc \times (\mu/\lqcd)^{d-1} \sim \Nc^2 
~\longrightarrow~
\mu \sim \Nc^{\frac{1}{d-1}} \lqcd \,,
\end{equation}
where we have multiplied color factors
for quarks, $\Nc$ and for gluons, $\Nc^2$.
In our world, $d=3$, so the gluon sector is modified
at $\mu \sim \Nc^{1/2} \lqcd$, as given in
Ref. \cite{McLerran:2007qj}.
This is parametrically larger than the scale, $\mu \sim \lqcd$,  where 
nuclear matter appears
and shortly later turns into quark matter \cite{McLerran:2007qj}.

This picture for deconfinement at finite density
implies that in spatial one dimension,
{\it if} the excitations are confined at zero density\footnote{
The analyses in this paper do not answer
whether excitations are confined at zero density.
In the large $\Nc$ limit, 
the theory is confining,
at least in the presence of arbitary 
small current quark mass \cite{'tHooft:1974hx}.
On the other hand, in the large flavor limit such that
$\Nc/\Nf \rightarrow 0$,
it is argued that the color sector takes a similar form
as the Schwinger model ---
Higgs phenomenon happens
due to the screening \cite{Armoni:1998ny}.},
they must survive even at very high density, 
$\mu \gg \lqcd$, because
the phase space for quark fluctuations 
remains the same as 
the vacuum case, and as a consequence, 
screening effects are not enhanced.
We will check that this is indeed the case
in our QCD$_2$ studies.

The second purpose of this paper
is to examine mechanisms of the  chiral symmetry breaking
near the Fermi surface
(modulo strong infrared phase fluctuations peculiar
to (1+1) dimensions 
\cite{KTtransition,Witten:1978qu,Coleman:1973ci}),
which recently attract much attentions in (3+1) dimensional
quark matter in the context of inhomogeneous chiral condensates
\cite{Kojo:2009ha,Kojo:2010fe,DGR1992,Son2000,Nickel:2009ke,Rapp:2000zd,Nakano:2004cd}.
We also study the relationship among quark density,
chiral condensates, and a scale generated in the colored sector.
In (1+1) dimensions,
such a relationship can be seen at the operator 
level \cite{Mandelstam:1975hb,Novikov:1982ei,Conformalbook,Tsvelik,Affleck:1985wb,Frishman:1992mr}.
Such operator relations do not strictly hold in higher dimensions.
But at high density such that 
the curvature of the Fermi surface is negligible, 
there emerge certain circumstances in which
low dimensional picture of
the Fermi surface is useful\footnote{
At high density,
transverse dynamics along the Fermi surface provides
negligible contributions, $\sim p_\perp^2/\mu$, to the energy spectra.
This can be used to factorize integral equations such as the
Schwinger-Dyson equation near the Fermi
surface by {\it integrating out} transverse dynamics,
instead of just ignoring it.
The resulting equation is dimensionally reduced one.
Such treatments are done in 
\cite{Kojo:2009ha,Kojo:2010fe,DGR1992,Son2000,Nickel:2009ke}.}.
The detailed understanding of QCD$_2$ would provide
a useful way of thinking for these situations.

One of the most enlightening approaches
to discuss the above issues in QCD$_2$ is the non-Abelian 
bosonization \cite{Novikov:1982ei}\footnote{
For other approaches based on the holography,
see Ref. \cite{Yee:2011yn}, for instance. }.
The use of it is motivated by at least two reasons.

First, the bosonized form allows us direct access
to the colorless objects such as quark number density or
chiral density.
A number of conclusions can be derived without explicitly
treating the colored objects and confining interactions among them,
which are not easy to deal with.
This is not the case in usual fermionic expressions
in which we have to built up color singlet quantities from
quark propagators.
The trouble is that even if we investigate a 
single quark propagator very precisely,
its property can not be directly converted into physical quantities
because residual confining interactions are so strong
\cite{Callan:1975ps}
--- the physical interpretation of results may be done only after
we construct colorless objects for which residual interactions are
under control.

The second reason comes from utility of 
the charge-color separation, analogous to 
the spin-charge separation in the condensed matter system
\cite{Tsvelik,Affleck:1985wb}.
In chiral limit, the quark number and 
colored sectors decouple as 
\cite{Affleck:1985wa,Frishman:1992mr}
\begin{equation}
 S_{{\rm fermion}} \longrightarrow 
S_{U(1)}[\varphi] + S_{{\rm color}}[h] \,,
\end{equation}
so that bosonized fields responsible
for $U(1)$ quark number ($\varphi$) 
and color densities ($h$)
can be treated separately.
It means that the color sector remains the same 
at any quark density, 
since quark chemical potential couples only to
quark number density, not to color density.
Hence confined excitations at zero density
persist up to arbitrarily high density.

The charge-color separation is also useful to 
get insights of the chiral symmetry near the Fermi surface.
In chiral limit, the chiral condensate in the bosonized form
has the following factorized structure,
\begin{equation}
 \la \bar{\psi}_L \psi_R \ra 
= \la \tr h \ra \la \rme^{\rmi \varphi} \ra \,.
\end{equation}
Since $\la \tr h \ra$ is unaffected by quark density,
density effects appear only through the exponent
$\la \rme^{\rmi \varphi} \ra$.
The color sector serves a massive scale
for the amplitude of the chiral condensate,
while the quark number sector provides a 
phase rotation.
Through the operator relations,
we will see a single baryon accompanies
a single chiral spiral.
Then it follows that
at nearly uniform quark density,
largely overlapped baryons,
which are merged with other baryons, 
induce a lot of chiral spirals with
period $\sim 1/2\mu$.
This period reflects that
the chiral condensates are made of co-moving
particles and holes near the Fermi surface, 
with momenta $p_f \sim \mu$.
Similar results have been found in other approaches
\cite{Schon:2000he,Bringoltz:2008iu}.
We will also argue that chiral symmetry breaking effects 
are relevant near the Fermi surface,
while not in most regions of the Fermi sea.

Throughtout our discussions,
we try to specify peculiarities in (1+1) dimensions
as much as possible.
Useful qualitative pictures applicable to (3+1) dimensions
may emerge only 
after the identification of such (1+1) dimensional specialities.

This paper is organized as follows.
In Sec.\ref{general},
we give general remarks on the Fermi sea
in Quarkyonic matter.
Then QCD$_2$ is briefly discussed in the 
fermionic language.
We will also explain why
we think of the quark Fermi sea, instead of 
the baryonic Fermi sea.
In Sec.\ref{1flavorvac},
the basics of the non-Abelian bosonization 
are quickly reviewed.
In Sec.\ref{baryon},
one baryon, two baryons, and finite baryon density
are pedagogically discussed.
The relationship between chiral spirals
and baryons or quark number density is shown.
In Sec.\ref{2flavordense},
we argue the extension to two flavor case.
In Sec.\ref{Impurities},
terms which enhance or suppress
the formation of chiral spirals are classified.
Sec.\ref{summary} is devoted to the summary
and discussions.
Throughout this paper,
we fix the value of $\Nc g_s^2$ 
when we change $\Nc$ from three.

\section{General remarks on the quark Fermi sea}
\label{general}
Before performing explicit calculations,
we argue qualitative aspects
of the Fermi sea in Quarkyonic matter.
To emphasize conceptual points,
we first consider very high quark density,
$\mu \gg \lqcd$, where
the inter quark distance is much smaller than
a typical distance scale of confinement.
Baryons already overlap one another
so that quarks need not belong to particular baryons.
The lower density will be discussed later
in order to emphasize how
nuclear and Quarkyonic matter are related.

Here we should specify our terminology
used in this paper.
The confinement and deconfinement 
will be classified by excitation modes.
We think that this will give more precise classifications
than those based on releasing of quarks from baryons.
For example, the former
distinguishes whether correlation functions for 
glueball opearators contain
deconfined mutli-gluon spectra or 
not\footnote{We have to admit that
practically it is not easy to make such a distinction 
in a solid way. 
For instance, in chiral limit and at zero density,
the spectral function for the $\rho$ meson channel 
contains
the cut of multi-pions starting from zero invariant mass.
On the other hand, we know that 
the height of multi-pion spectra of $O(\Nc^0)$  is much lower than 
that expected from $q\bar{q}$ continuum of $O(\Nc)$.
So we can interpret this hierarchy as a consequence of confinement.
Unfortunately this classification becomes again ambiguous
in the quark-hadron duality region at high energy \cite{Poggio:1975af}.
}.

What are relationships between
asymptotic freedom and quark density?
When the inter quark distance is small,
typical interactions are supposed to be
weak because of asymptotic freedom.
This logic is frequently used to justify
weak coupling treatments of high density QCD
as well as to guarantee
the picture of the degenerated quark Fermi sea.

\begin{figure}[tb]
\begin{center}
\scalebox{1.0}[1.0] {
  \includegraphics[scale=.15]{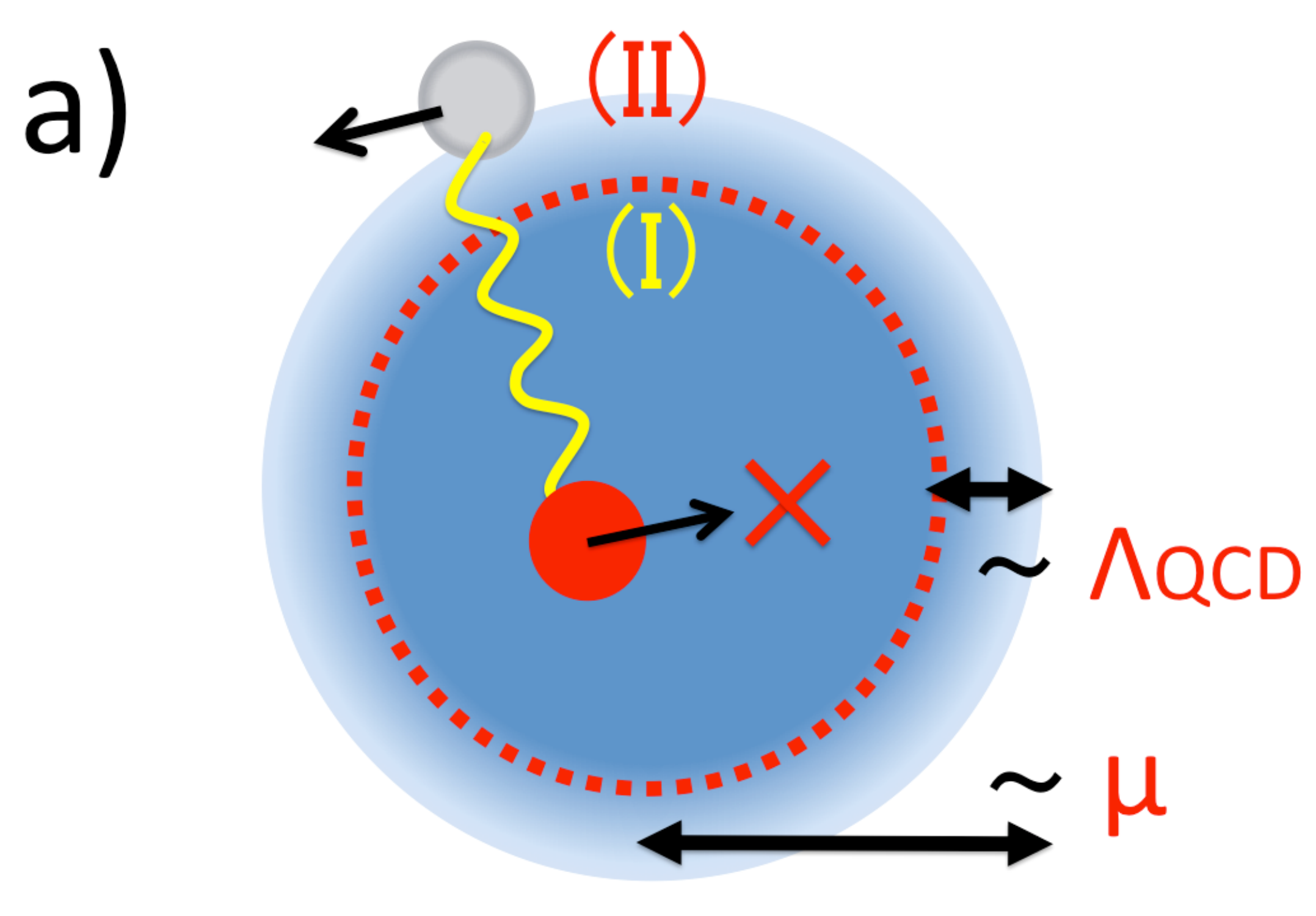} }
\hspace{0.8cm}
\scalebox{1.0}[1.0] {
  \includegraphics[scale=.15]{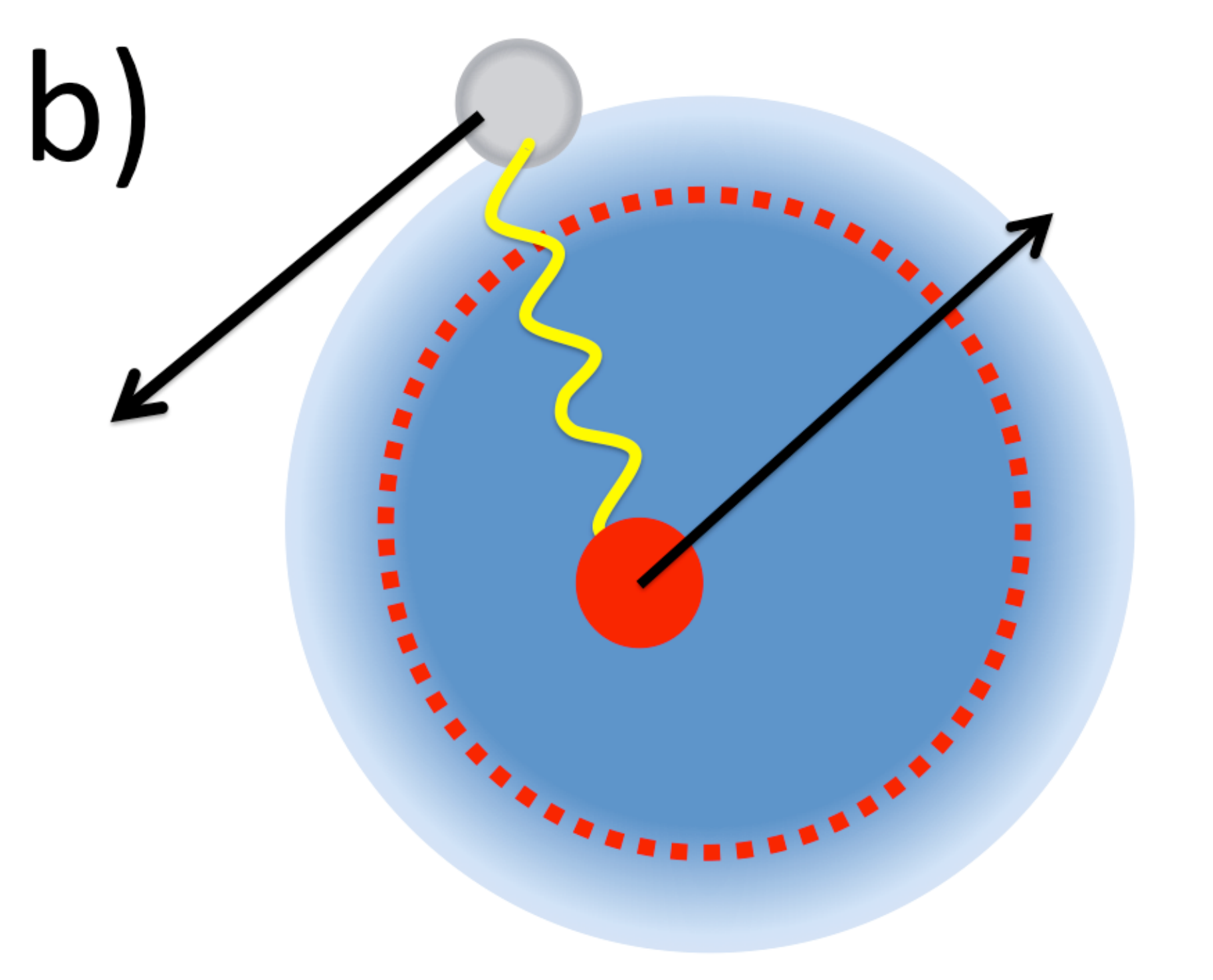} }
\end{center}
\vspace{0.2cm}
\caption{The scattering of quarks in the region (I), 
$|\vec{p}| \lesssim \mu-\lqcd$.
(a) Small momentum exchange processes,
which are forbidden by Pauli blocking.
(b) Hard momentum exchange processes,
which transfer quarks from the region (I) to its outside.
(Dotted lines separate the regions (I) and (II).)
}
\label{fermi1}
\vspace{0.2cm}
\end{figure}

For more precise statements,
we divide regions of the Fermi sea
into (I) an inner region ($|\vec{p}| \lesssim \mu - \lqcd$)
and (II) a surface region ($|\vec{p}| \gtrsim \mu - \lqcd$)
(see Fig.\ref{fermi1}(a)).
Let us clarify properties
of interacting quarks in each region,
by arguing possible corrections
from virtual scattering processes.

In the region (I), Pauli principle
prevents quarks from being scattered by 
small momentum exchanges
(Fig.\ref{fermi1}.(a)).
Thus quarks in the region (I) have little chance
to feel nonperturbative effects.
Allowed processes are hard scatterings
which transfer quarks to the outside of the Fermi sea
(Fig.\ref{fermi1}.(b)).
The latter may be treated within weak coupling methods. 
Therefore quarks in the region (I) can be described
by quasi-particles affected by perturbative
self-interactions and many body effects.

\begin{figure}[tb]
\begin{center}
\scalebox{1.0}[1.0] {
  \includegraphics[scale=.25]{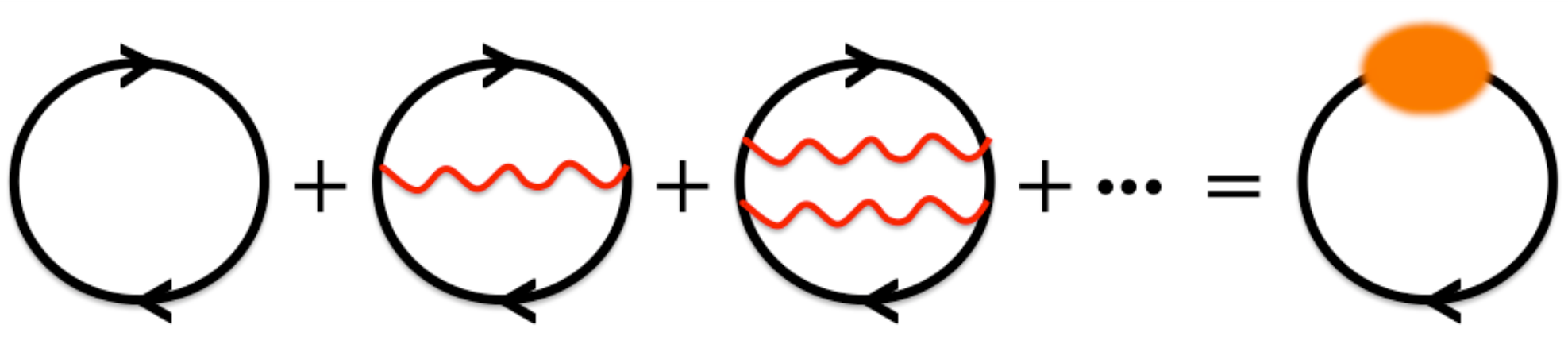} 
}
\caption{A uncorrelated sum of the single quark loops.
The Fock term is a part of the full quark propagator
at large $\Nc$.}
\label{Qloop}
\end{center}
\vspace{0.1cm}
\end{figure}

The above argument is incomplete, 
since there are also contributions from
zero momentum exchange processes,
in which quarks just exchange positions
in the Fermi sea 
(the Fock term, the second diagram in Fig.\ref{Qloop}).
Thereby those processes
are allowed inside of the Fermi sea,
and should be sensitive to
the gluon propagator at zero momentum.
It would cause problems in computations
of several quantities such as the total free energy.
Such computations include quark loops
which may contain virtual gluon loops with
zero momentum transfer.
For some gauge, gluon propagator shows divergent behavior
in the deep IR region,
so such diagrams would cause the IR divergence\footnote{
A similar problem has been discussed in case of an electron gas,
and usually one handles this by including the Debye screening mass
to kill the infrared divergences from photon propagators.}.

\begin{figure}[tb]
\begin{center}
\scalebox{1.0}[1.0] {
  \includegraphics[scale=.25]{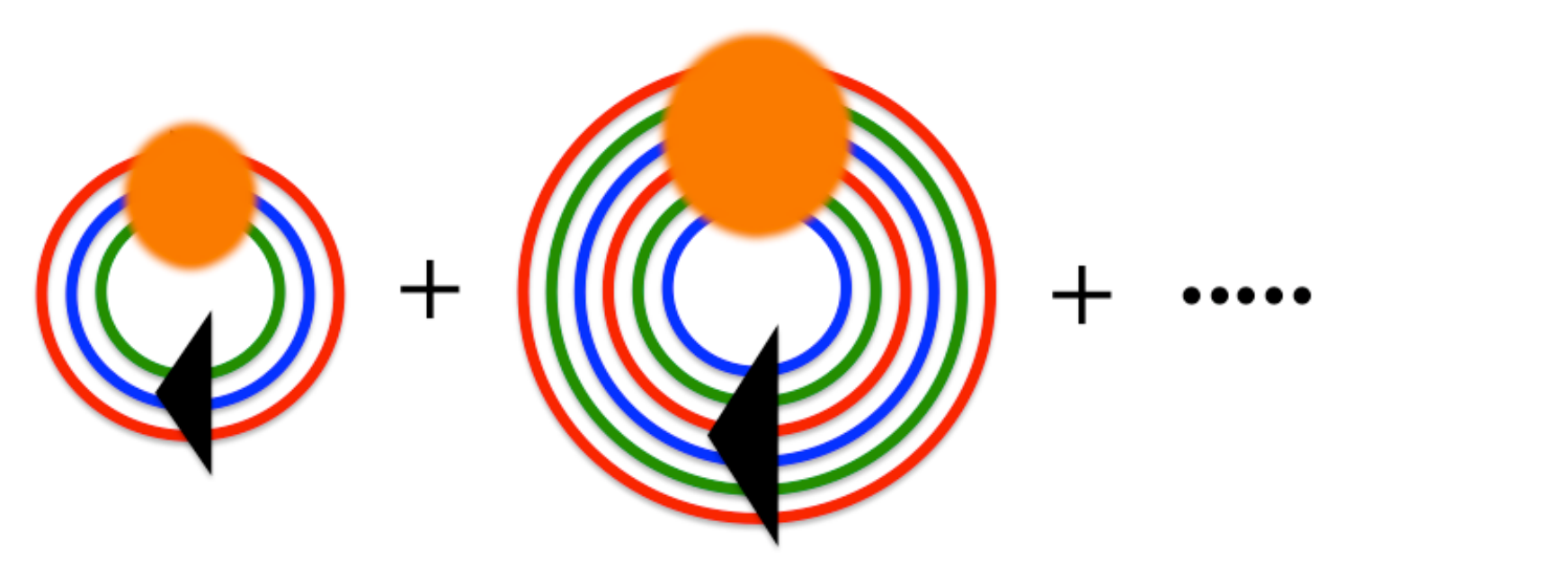} 
}
\caption{A sum of the color singlet loops with $n\times \Nc$ quarks,
where $n$ is an integer.
The $\Nc=3$ case is shown.
In the dilute regime, these diagrams can be
factorized into a sum of baryon loops with a few number
of the meson interactions.
In the dense regime, the quark exchange frequently occurs,
so we need more general descriptions than the
baryonic matter.
}
\label{Bloop}
\end{center}
\vspace{0.1cm}
\end{figure}

If we compute the total free energy
as an {\it uncorrelated} sum of the {\it single} 
quark loops
(Fig.\ref{Qloop}),
the deep IR behavior of gluon propagators 
is a certainly serious problem.
However, in computations of the color singlet Fermi sea,
it may be possible that
the deep IR contributions
cancel out after self-consistent treatments or resummation
together with other quark loops.
Such contributions can not be mimicked by
an uncorrelated sum of single quark loops ---
we cannot derive any physical conclusions
for color singlet objects
until we correctly take into account the interactions among quark loops
which make diagrams color singlet (Fig.\ref{Bloop}).

There is a case study for the $1/\vec{p}^4$
confining propagator in (3+1) dimensions.
It indicates that the deep IR contributions
at $\vec{p}=\vec{0}$ are canceled out after
consistent treatments of the quark self-energy and
confining interactions in the color singlet 
objects\footnote{The simplest way
to show the cancellation is to just use 
the principal value regulator for the $1/\vec{p}^4$ force
for which the propagator takes zero value at 
$\vec{p}=\vec{0}$ \cite{Callan:1975ps}.
In Ref.\cite{Glozman:2008fk}, usual IR cutoff scheme is used 
to argue the cancellation.
While two schemes provide different quark propagators,
they give the same spectra at the level of the Bethe-Salpeter
equation.
Thus the detail of $\vec{p}=\vec{0}$ is irrelevant
as far as consistent calculations are performed.}.
Intuitively, this cancellation may reflect that
the gluon propagator at $\vec{p}=\vec{0}$, 
which corresponds to an infinitely long string 
configuration\footnote{In the usual IR cutoff scheme,
$- \int \rmd\vec{p} ~
\frac{\sigma}{ (\vec{p}^2 + \mu_R^2)^2 } \rme^{\rmi \vec{p}\cdot \vec{r} }
\sim - \frac{\sigma}{\mu_R} \rme^{-\mu_R r}
\sim - \frac{\sigma}{\mu_R} + \sigma r + O(\mu_R r^2)$,
where $\mu_R$ is infinitesimal.
The first term is independent of $\vec{p}$,
and expresses a string with infinite length.
},
can not be relevant inside of color singlet objects.

Although we are not aware of whether 
this cancellation generically holds 
in any confining models in (3+1) dimensions,
we will rely on this qualitative interpretation 
for the nonperturbative part.
In (1+1) dimensions, we will see that
no problems arise from the confining force at finite density.

\begin{figure}[tb]
\begin{center}
\scalebox{1.0}[1.0] {
  \includegraphics[scale=.15]{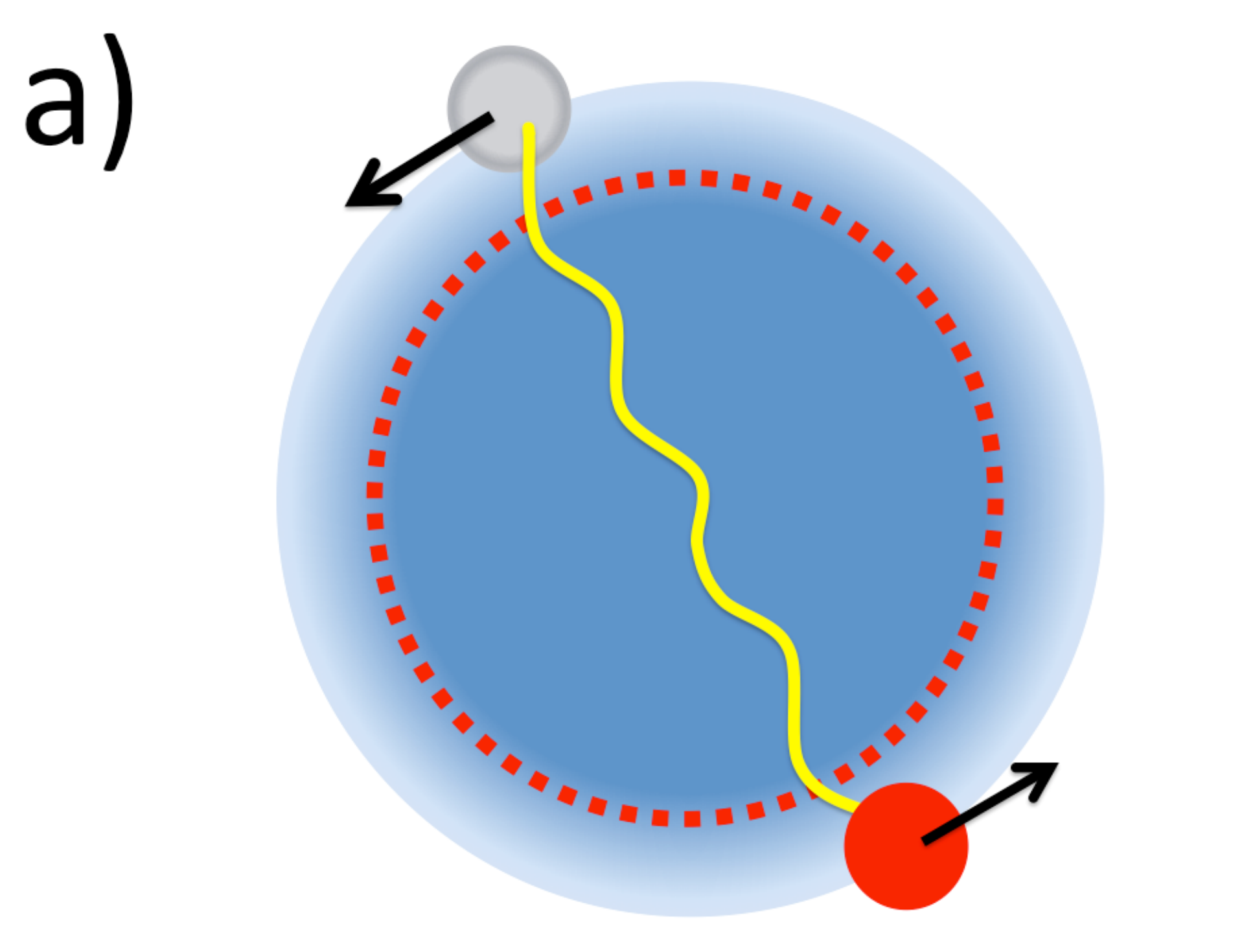} }
\hspace{1.1cm}
\scalebox{1.0}[1.0] {
  \includegraphics[scale=.15]{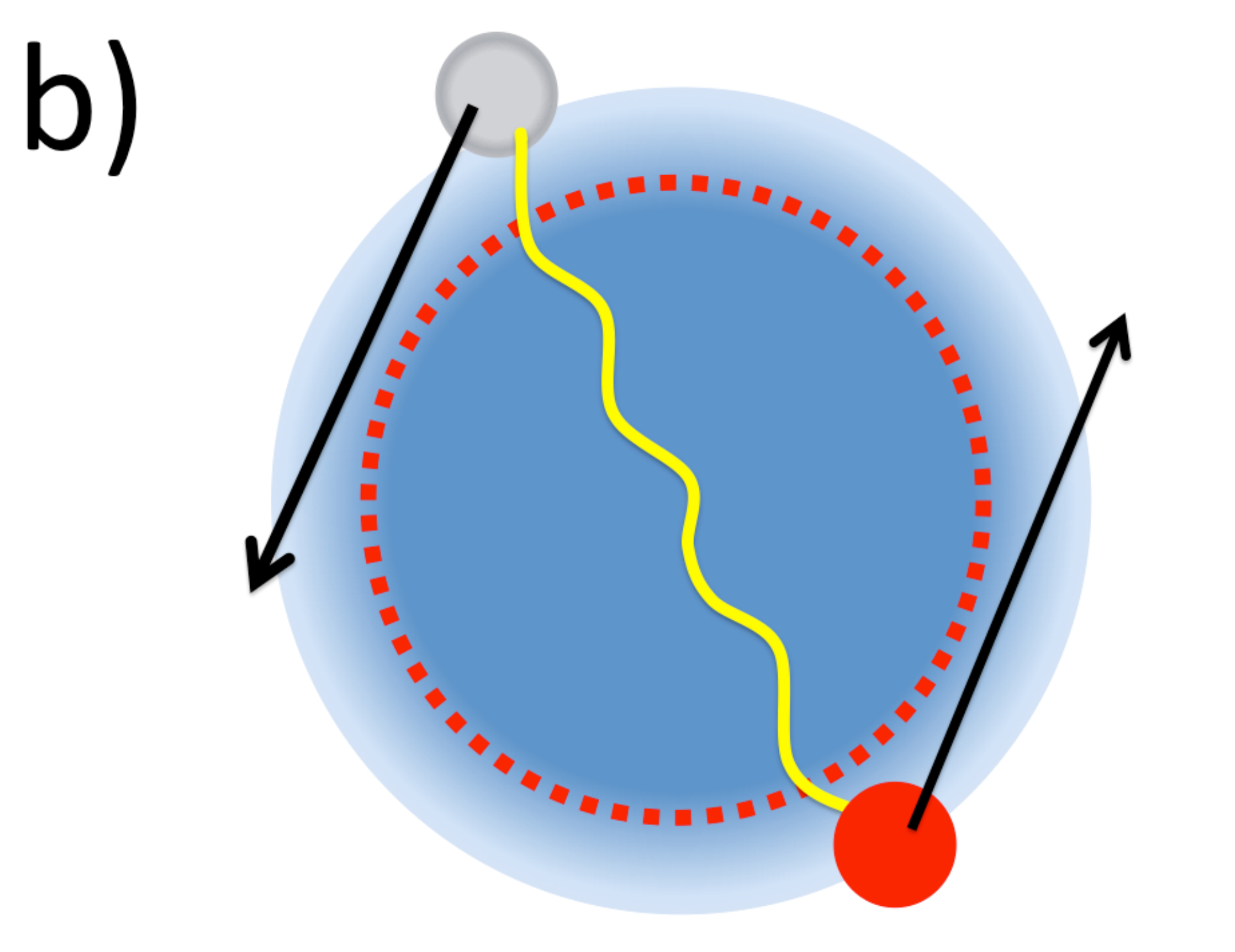} }
\end{center}
\caption{The scattering of quarks in the region (II), 
$|\vec{p}| \gtrsim \mu-\lqcd$.
(a) Small momentum exchange processes
as sources of nonperturbative phenomena.
(b) Hard momentum exchange processes.
Both processes are allowed.}
\label{fermi2}
\vspace{0.2cm}
\end{figure}

Now let us consider the region (II).
There soft momentum exchanges 
(Fig.\ref{fermi2}.(a))
are allowed in the region (II).
In addition to perturbative effects 
(Fig.\ref{fermi2}.(b)),
nonperturbative effects on quarks also operate here,
and properties of quarks near the Fermi surface 
may be strongly modified from
those of quasi-free fermions.

\begin{figure}[tb]
\begin{center}
\hspace{0.1cm}
\scalebox{1.0}[1.0] {
  \includegraphics[scale=.30]{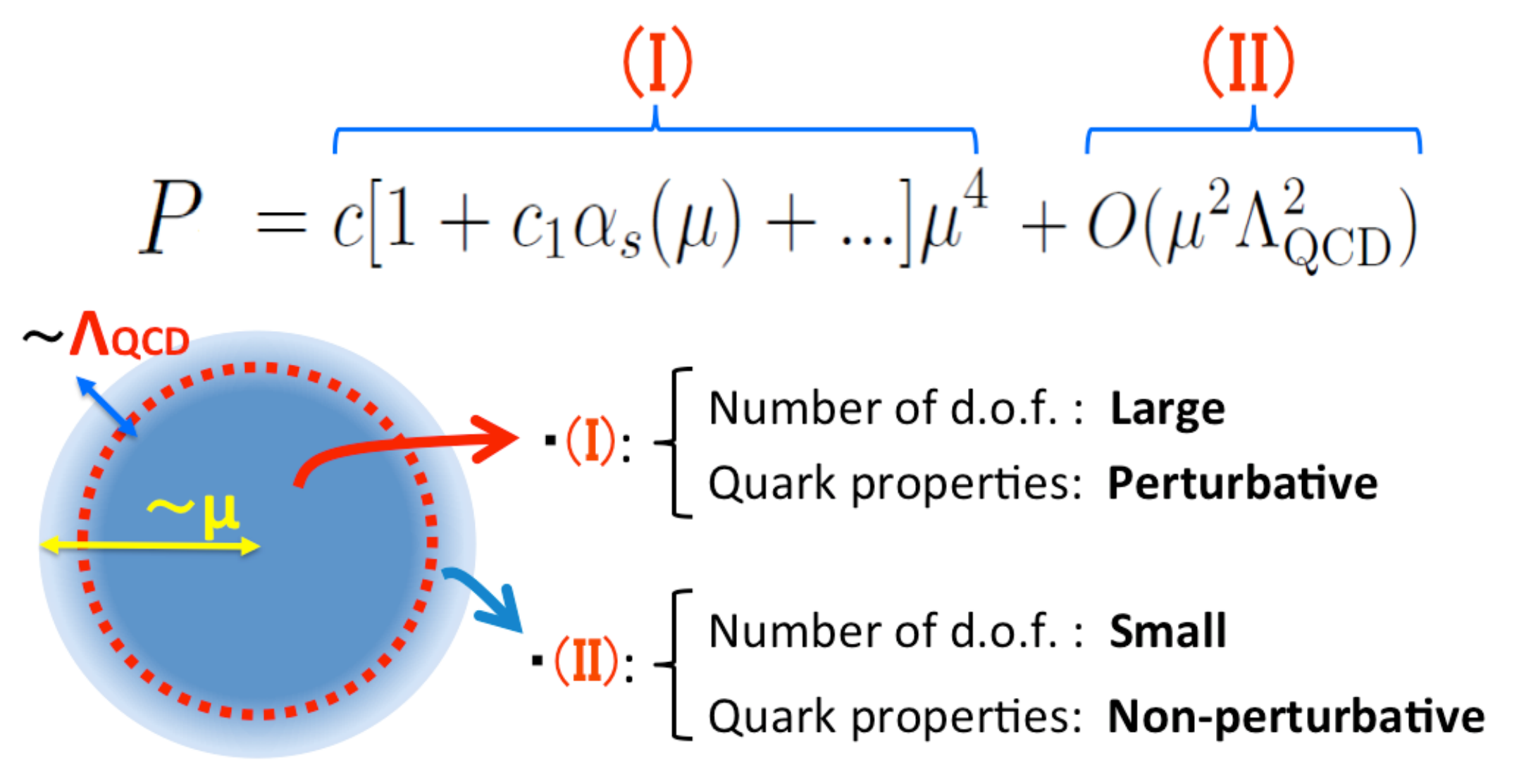} }
\caption{The pressure as an example of bulk quantities.
The dominant contribution ($\sim \mu^4$) 
comes from the region (I) including
a large number of (perturbative) quarks.}
\label{pressure1}
\vspace{0.2cm}
\end{center}
\end{figure}

With these pictures for quarks,
let us first examine bulk thermodynamic quantities 
to which all quarks in the Fermi sea contribute.
A schematic picture is given in Fig.\ref{pressure1},
taking the pressure as an example.
At high density,
they are well saturated by contributions from the region (I),
simply because a number of quarks is much larger
than those in the region (II).
The region (I) gives contribution of $\sim \mu^4$
including perturbative corrections,
while we get small (nonperturbative) 
contribution of $\sim \mu^2 \lqcd^2$
from the region (II).
In this sense, 
the picture of perturbative quarks should work
in computations of bulk quantities,
and should reproduce perturbative results \cite{Freedman:1976xs}.

Again, we do {\it not} insist
that the pressure in Quarkyonic matter can be computed
from a sum of non-interacting single quark loops
in a literally sense.
Interacting diagrams, which include other quark lines in a
color singlet way,
must be included to avoid the problem 
of infinitely long strings generated from quarks.
Our claim is that once this particular infrared problem is
handled, quarks inside of the Fermi sea should look like
perturbative quarks.

While the bulk quantities should be well-approximated
by perturbative contributions,
physics near the Fermi surface is
sensitive to the excitation modes and nonperturbative effects.
Examples are the phase structures,
transport phenomena, etc.
The issues of confinement and chiral symmetry breaking
are nontrivial here.

To examine pictures discussed so far,
let us take one flavor masselss \qcd as an example.
The action of \qcd with axial gauge fixing is
\begin{equation}
S = \int \rmd^2x \,
 \bar{\psi}(x) (\rmi \Slash{\partial} + \mu \gamma^0) \psi (x)
+ \int \rmd^2x \rmd^2y
\, J_\rmA^\mu (x) D^{\rmA \rmB}_{\mu \nu}(x-y) J_\rmB^\nu (y) \,,
\end{equation}
where $x=(t,z)$, $J_\rmA^\mu = \bar{\psi}\, t_\rmA \gamma^\mu \psi$,
and $t_\rmA$ is usual color matrix in the foundamental
representation. Its normalization is 
$\tr[t_\rmA t_\rmB]=\delta_{\rmA \rmB}/2$.
The interactions between
colored currents are confining, and instantaneous in axial gauge,
$D_{\mu \nu}^{\rmA \rmB}(x-y) 
= \delta^{\rmA \rmB} \delta_{\mu 0} \delta_{\nu 0} |\vec{x}-\vec{y}|$.
The gluon propagator
is not dynamical in (1+1) dimensions, 
so we have eliminated the gluon fields using 
the equation of motion.
Note that the gauge coupling constant
has a dimension one, and 
serves the scale $\lqcd$ in (1+1) dimensions.

In the presence of the Fermi sea,
it is convenient to redefine quark fields
depending on their moving directions $+z$ or $-z$,
\begin{equation}
\psi_\pm(t,z) = \rme^{\pm \rmi\mu z} \psi_\pm'(t,z) \,.
\label{shift}
\end{equation}
Its shorthand notation is 
$\psi = e^{\rmi \mu \gamma_5 z}\psi'$, 
since eigenvalues of $\gamma_5 = \gamma_0 \gamma_z$
coincide with moving directions in (1+1) dimensions,
\begin{equation}
\bar{\psi} \rmi \Slash{\partial} \psi
= \psi_-^\dag \rmi (\partial_0 + \partial_z) \psi_-
+ \psi_+^\dag \rmi (\partial_0 - \partial_z) \psi_+ \,.
\end{equation}
So Eq.(\ref{shift}) just means shifts of momenta to
measure them from the Fermi surface,
$\pm \mu$.
With these new fields, we obtain
\begin{equation}
 \bar{\psi}(x) (\rmi \Slash{\partial} + \mu \gamma^0) \psi (x)
~\rightarrow~ \bar{\psi}'(x) \rmi \Slash{\partial} \psi' (x)
\, , ~~ 
J_\rmA^{\mu} (x) ~\rightarrow~ (J'_\rmA)^\mu (x) \,,
\end{equation}
and the Lagrangian becomes that of the zero density.

Since Lagrangian is the same as that in vacuum, 
properties of excitations such as energies
are also unchanged except their shifted momenta.
Therefore in (1+1) dimensions,
if models have confined excitations in vacuum,
they also do at finite density.
This is true no matter how quark density is high
and inter quark distances are short.
The value of $\Nc$ is not essential 
in the present discussion.
An overlap of baryons and
deconfinement do not have one to one correspondence. 

In coordinate space,
this situation may be difficult to imagine
if we view quarks as 
point-like particles as classical mechanics.
Perhaps it is easier to understand
if we see the distribution of colors
by focusing on the wave properties of quarks.
Closely packed quarks form a color singlet background
minimizing the color charge in the system,
and excitations are deviations from such a background.
Quarks and quark-holes always appear together
as confined excitations.

To what extent can we generalize these pictures
to the higher dimensions?
The main changes can be found in the phase space
allowed for colored $q \bar{q}$ fluctuations
which screen the exchange of gluons.

In (1+1) dimensions, phase space near the edge of
occupied states are always the same in the Dirac and Fermi sea.
Thus the strength always looks same in a whole density region.
This explains why the excitation energy near the Fermi surface
does not change even at asymptotically high density.

In spatial dimensions larger than one,
the phase space for fluctuation modes increases
as density does.
Eventually
interactions between colors are strongly screened,
or squeezed color fluxes dissociate completely.
Then weak coupling methods become enough to 
describe not only bulk quantities but also excitation modes.

Currently we have only a rough estimate
of such density for the (3+1) dimensional system, 
$\mu \sim \Nc^{1/2} \lqcd$,
paramerically larger than density where baryons overlap,
$\mu \sim \lqcd$.
The relevance of the present discussions to (3+1) dimensions
depends on how widely these two scales are separated
after including all other numerical factors.
Attempts to estimate such a width can be found
in several model studies 
\cite{Herbst:2010rf,Fukushima:2010is,Lottini:2011zp}.

So far we have considered only high density regime
in which the concept of Quarkyonic matter
is most clearly illustrated.
Now let us consider how such high density regime 
is connected to lower density regime where
the presence of nuclear matter is important.

Perhaps it is good to start with
seeing how the picture of the baryonic Fermi sea 
in nuclear matter breaks down
as density increases.
The main problem is the difficulty to maintain
the quasi-particle picture of baryons.
Typical interactions among baryons
get stronger at shorter distance,
or in harder momentum transfer processes.
This means that baryons near
the Fermi surface and deeply inside of the Fermi sea
can strongly affect one another.
This badly destroys the concept of the baryonic Fermi sea 
at very high nuclear density\footnote{
The reconstruction of the Fermi sea should happen at some density
without depending on whether $\Nc$ is odd or even,
or whether baryons are composite fermions or bosons.
Even if baryons are composite bosons,
we start to observe their internal structures made of fermions 
at density of the baryon overlap ---
the Pauli principle acts on the internal momenta of fermions.
}.

After all, at high density, we have no right to stick to 
the Fock space made of conventional baryons made of $\Nc$ quarks.
The Fock space including baryonic objects made of $2\Nc, 3\Nc, \cdots$ 
quarks, which are also color singlet,
may become equally important (Fig. \ref{Bloop}).
This is so, because energetically such states 
may be easily 
reached by strong forces among ordinary baryons as well as 
by large collision rates at high baryon density.
We will give more detailed arguments
in Sec. \ref{1flavordense}.

This signals that effective degrees of freedom
should be switched from baryons to quarks,
in order to temper the growth of residual forces 
among the objects which we chose
as basic degrees of freedom.
At shorter distance, most of quarks feel weaker interactions,
so they are reasonable alternatives forming the Fermi sea.
Hence we expect that the transition from nuclear to Quarkyonic matter
smoothly proceeds inside of the Fermi sea,
while in both matters excitations remain confined.

Here we should stress again that
the basic ingredients of the above arguments
are properties of interactions --
whether composite objects overlap or not,
is the secondary issue.
For instance, if the baryon based descriptions
gave smaller corrections of interactions
than those in quark descriptions,
we could continue to use
the picture of the baryonic Fermi sea
even at very high density.
Of course such a situation is very unlikely in QCD.
Needless to say, 
knowledges about baryon-baryon interactions at high density
are crucial for further arguments.

In summary,
Quarkyonic matter differs from conventional quark matter
in the excitation modes,
and should be distinguished from nuclear matter
by bulk quantities.

\section{The non-Abelian bosonization of QCD$_2$ (One flavor)}
\label{1flavorvac}
From this section, 
we start computations using the non-Abelian bosonization.
The utility of the bosonized form is 
that we can compute several quantities in terms of
quark number and color densities,
emphasizing how differently
a quark chemical potential acts 
on quark numbers and colors.
Another utility is that we can write
explicit relations between
chiral and quark number densities
which are useful to consider issues of the
chiral symmetry breaking/restoration in quark matter.

\subsection{Preliminaries}

We review the non-Abelian bosonization rules,
to the extent necessary for our arguments
(For quick introduction, see \cite{Affleck:1985wb}).
The rules take compact forms by using lightcone coordinates
$x_\pm = x^\mp/2 = x_0 \pm x_1 $, and
$\partial_\pm = 2 \partial^\mp = \partial_0 \pm \partial_1$.
The currents for $U(1)$ and $SU(\Nc)$ chiral charges
are
\begin{align}
&J_- = \rmi\, \frac{\Nc}{4\pi} U \partial_- U^\dag
= \ :\! \psi^\dag_- \psi_- \!: \, ,
~~
J_+ = \rmi\, \frac{\Nc}{4\pi} U^\dag \partial_+ U
= \ :\! \psi^\dag_+ \psi_+ \!: \, ,
\nonumber \\ 
& \hspace{-0.5cm}
J^\rmA_-  = \frac{\rmi}{2\pi} \tr[ h \partial_- h^\dag t_\rmA]
= \ :\! \psi^\dag_- t_\rmA \psi_- \!: \, ,
~~
J^\rmA_+  = \frac{\rmi}{2\pi} \tr[ h^\dag \partial_+ h t_\rmA]
= \ :\! \psi^\dag_+ t_\rmA \psi_+ \!: \, ,
\end{align}
with fields $U(x)$ and $h(x)$
which are frequently written as 
$U(x) = e^{ \rmi \phi(x)}$ and $h(x) = e^{ \rmi t_\rmA \pi_\rmA(x)}$.
We have normal ordered currents and
subtracted infinite constants.
These currents are related to $J_\mu$ and $J_{\mu 5}$ as
\begin{equation}
J_\mu = 
\left(\begin{array}{ccc}
 J_0  \\
 J_1
\end{array}
\right)
=
\left(\begin{array}{ccc}
 J_- + J_+  \\
 J_+ - J_-
\end{array}
\right)~,
\ \ \ 
J_{\mu 5} = \epsilon_{\mu \nu} J^\nu
=
\left(\begin{array}{ccc}
 J_- - J_+  \\
 J_- + J_+
\end{array}
\right)~.
\end{equation}
In chiral limit, they satisfy 
$\partial^\mu J_\mu = \partial^\mu J_{\mu5}
= \partial_\pm J_\mp = 0$.
Similar relations hold for $J^\rmA$.
The operator responsible for the chiral density is
\begin{equation}
\bar{\psi}^a_+ \psi^b_- 
= c(M)~ (h^{ab})_M ( U)_M \, ,
\end{equation}
where $a,b$ is color indices of fermions.
Here $c(M)$ is a renormalization constant,
and a subscript $M$ means that operators are
normal ordered at a scale $M$ \cite{Coleman:1974bu}.
Note that the chiral density and
the quark number density are
characterized by common fields $U$ and $h$,
so once we know behaviors of these fields,
we also know both densities simultaneously.

The kinetic part of the fermion 
becomes
\begin{equation}
 \int \rmd^2x~
 \bar{\psi}(x) \rmi \Slash{\partial} \psi (x) \ 
\longrightarrow \ 
 S^{ {\rm U(1)} }_{k=\Nc} [U] + S_{k=1}^{{\rm WZW}}[h] \,,
\end{equation}
where $S_{k=\Nc}^{ {\rm U(1)} }[U]$ is an action of free massless boson,
\begin{equation}
S^{ {\rm U(1)} }_{k=\Nc} [U]
 =  
\frac{\Nc}{8\pi} \int \rmd^2 x \, 
(\partial_\mu U^{\dag}\partial^\mu U) \,,
\end{equation}
and $S^{{\rm WZW}}_{k=1}[h]$ is
Wess-Zumino-Novikov-Witten (WZNW) action
\cite{Novikov:1982ei},
\begin{equation}
S_{k}^{{\rm WZW}}[l]
 = \frac{k}{8\pi} \tr \bigg[ 
 \int \rmd^2x\, \partial_\mu^\dag l \partial^\mu l 
+ \frac{2}{3} 
 \int \rmd^3x\, \epsilon^{\mu \nu \lambda} 
 (l^{\dag} \partial_\mu l) (l^{\dag}\partial_\nu l) 
   (l^{\dag} \partial_\lambda l) 
\bigg] \, ,
\end{equation}
with level $k=1$, and $l=h$.
While the $\beta$ function of 
the non-linear $\sigma$ model shows an asymptotic free
behavior, the WZW term cancels out it.
Thus the action is conformal at quantum level.
The scale $\lqcd$ will be introduced by 
confining current-current interactions
($g\sim \lqcd$).

Now we found one of the utilities of
the non-Abelian bosonization.
The colored currents include only
colored boson fields $h(x)$,
and do not depend on other densities in 
chiral limit:
The actions for $U(1)$ quark number 
and color charge sectors decouple.
Only the color sector includes the dimensionful
quantity, i.e., the gauge coupling constant
with dimension one. 
The decoupled $U(1)$ action is 
described by the free bosons and is conformal.

\subsection{The infrared fluctuations
and the spontaneous symmetry breaking}
Before proceeding to our main discussions,
first let us specify peculiarities in (1+1) dimensions.
We will shortly take a glance at the relationship between
the conformal property in the $U(1)$ action 
and Coleman-Mermin-Wagner theorem:
the spontaneous symmetry breaking
is forbidden in (1+1) dimensions
\cite{Coleman:1973ci}.

Because of the separation of the $U(1)$ and color
sectors,
the chiral condensate can be written in the 
factorized form,
\begin{equation}
\la \bar{\psi}_+ \psi_- \ra
 = c(M) \la \tr [h]_M \ra \la U_M \ra \, .
\label{condensate}
\end{equation}
This matrix element vanishes due to
the infrared divergence in the propagator of 
the massless $U(1)$ boson
which characterize the phase degrees of freedom
in the chiral space. 
Intuitively, this reflects that phase fields
in the chiral space rotate rapidly,
without taking any particular direction.
An explicit expression is ($\beta = (4\pi/\Nc)^{1/2}$)
\cite{Coleman:1974bu}
\begin{equation}
\hspace{-0.4cm}
\la U_M \ra 
= \la ( \rme^{ \rmi \beta \varphi(x;m_\varphi)})_M \ra
= \rme^{ - \frac{\beta^2}{2} \Delta(M;m_\varphi)}
= \rme^{ \frac{\beta^2}{8\pi} \ln\frac{m_\varphi^2}{M^2} }
=
\bigg( \frac{m_\varphi }{M} \bigg)^{ 1/ \Nc} \, .
\end{equation}
Witten first observed this sort of
$\Nc$ dependence in (1+1) dimensional correlation functions
of chiral operators, which behave as
$\sim 1/|\vec{x}-\vec{y}|^{1/\Nc}$ \cite{Witten:1978qu}.
Here $\Delta$ is a propagator of the boson $\varphi$,
and $m_\varphi$ is the mass of $\varphi$
which should be taken to be zero at the 
end of the calculations.

This expression is one example which 
illustrates subtleties in the limiting order, 
large $\Nc$ and chiral limit.
If we start with the strict chiral limit,
we must first take $m_\varphi \rightarrow 0$ limit
before taking the large $\Nc$ limit.
Thus the matrix element vanishes.

The exception is the case with finite volume or
with the infrared momentum cutoff.
This is a typical situation studied in seminal works.
They essentially yield similar effects to the 
introduction of $m_\varphi$.
Once we regularize these infrared fluctuations,
we can simply take the large $\Nc$ limit,
\begin{equation}
\la \bar{\psi}_+ \psi_- \ra
 = c(M) \la \tr [h]_M \ra 
\times \bigg( \frac{m_\varphi }{M} \bigg)^{ 1/ \Nc}
\longrightarrow ~
c(M) \la \tr [h]_M \ra \, .
\label{condensate2}
\end{equation}
Here the $m_\varphi$ dependence disappears
as $\Nc \rightarrow \infty$.
The renormalization 
constant and the value of $\la \tr [h]_M \ra$
will be fixed in the next subsection. 
In the following we will frequently
omit a subscript $M$ if it does not play
an essential role.

Finally we would like to recall
Witten's argument \cite{Witten:1978qu} that
phase fields in (1+1) dimensions
should not be identified as Goldstone bosons.
While both of them share common features
as phase fluctuations, 
they play quite different roles in the chiral limit.
The former belongs to the ground state properties,
while the latter appears as an excitation from the vacuum.
This aspect becomes a little bit vague
once we introduced an explicit breaking of the
chiral symmetry,
since phase fluctuations become physical excitation modes
after the vacuum chose a particular direction in the chiral space.
In what follows,
we will not argue issues which are very sensitive to 
this conceptual issues.
We will regard phase fluctuations as excitation modes.

\subsection{A mass perturbation}
Now let us introduce a current quark mass term
\begin{equation}
{\mathcal L}_m 
= - m_q (\bar{\psi}_+ \psi_- + \bar{\psi}_- \psi_+ )
 = - m_q \times c(M) ( \tr [h] U 
+ \tr[h^\dag] U^\dag ) \, .
\end{equation}
which explicitly breaks the conformal symmetry 
in the $U(1)$ sector.
Besides the infrared regularization,
the mass term couples the $U(1)$ and color sectors,
thus it is no longer possible to 
investigate these sectors separately.
Thus from now we rely on the large $\Nc$ limit.
Then we can apply a probe approximation
which significantly simplifies the arguments
\cite{Affleck:1985wa}.

The point is that while 
the $U(1)$ sector provides $O(\Nc)$ contributions at most,
those from quantum fluctuations in the color sector
is $O(\Nc^2)$ because of $O(\Nc^2)$ degrees of freedom.
The confining interactions generate
a mass gap for the colored particles,
so we can replace $\tr h$ with $\la \tr h \ra$ in arguments of
the low energy phenomena.
The value of $\la \tr h \ra$ is supposed to be well-approximated
by that computed without a current quark mass
which is expected to act as a small perturbation to the mass gap
$\sim \lqcd$.

We analyze the $U(1)$ sector
under such a background.
Then the effective action for the low energy excitations is
\begin{equation}
S^{\eff}[U]
 \simeq \frac{\Nc}{8\pi} \int \rmd^2 x 
\big[ (\partial_\mu U^{\dag}\partial^\mu U)
 + m_\varphi^2 (U + U^\dag -2 ) \big] \, ,
\end{equation}
where we have subtracted a constant to normalize
the action.
We have also used $\la \tr h \ra = \la \tr h^\dag \ra$
and defined 
$m_\varphi^2 = - m_q \times 8\pi c(M) \la \tr [h] \ra/\Nc$
for later convenience.
Of course we can arrive at the similar Lagrangian 
following usual steps, the bottom up construction
with the chiral symmetry constraints.
Only difference is that in the above treatment 
the role of the colored sector was explicit.

Now we take the expression 
$U= \rme^{ \rmi (4\pi/\Nc)^{1/2} \varphi}$
to canonically normalize
the kinetic term for quantum excitations.
Then the action for $\varphi$ becomes
\begin{align}
S^{\eff}[\varphi]
 &= \int \rmd^2 x \left(
 \frac{1}{2} (\partial_\mu \varphi)^2
 + \frac{\Nc m_\varphi^2}{4\pi} 
\big( \cos(\sqrt{4\pi/\Nc} \varphi ) -1\big) \right) 
\nonumber \\
&= \int \rmd^2 x \, 
 \frac{1}{2} \left( (\partial_\mu \varphi)^2
 - m_\varphi^2 \varphi^2
\right) + O(1/\Nc) \, .
\end{align}
At large $\Nc$, 
$m_\varphi$ should coincide with the mass of the lowest mode 
of the Bethe-Salpeter equations, computed by 't Hooft. 
He computed the pole mass of the channel $J_-$
in the lightcone gauge. 
In the bosonized form, the long range behavior
of the corresponding correlator is ($x \gg (g^2\Nc)^{-1/2}$)
\begin{equation}
\la J_- (x) J^\dag_-(0) \ra
\sim \la \partial_- \varphi(x) \partial_- \varphi(0) \ra \, ,
\end{equation}
so the pole should coincides with $m_\varphi^2$.
Matching 't Hooft's result with that in the bosonized form,
\begin{equation}
m_\varphi^2 = m_q \sqrt{ \frac{4g^2 \Nc \pi }{3}}
= - m_q  \frac{8\pi c(M) \la \tr [h] \ra }{\Nc} \, ,
\end{equation}
then we can fix the scale, $c(M) \la \tr[h] \ra$.
Using these constants, 
we can see the value of the chiral condensate.
Since $m_\varphi$ regulates infrared behaviors of $\varphi$,
the $U(1)$ sector now has a nonvanishing expectation value.
At large $\Nc$, we have
\begin{equation}
\la \bar{\psi}\psi \ra 
= 2 c(M) \la \tr [h] \ra 
= - \Nc \sqrt{ \frac{g^2 \Nc}{12\pi} } \, .
\end{equation}
This result was originally derived 
by Zhitnitsky using the current algebra plus
the operator product expansion
\cite{Zhitnitsky:1985um}.
We reproduce this result simply because
the bosonization procedures correctly implement
the approximate chiral symmetry,
thus satisfy the current algebra aspects.
For later convenience, we write
the pion mass in terms of the chiral condensate
\begin{equation}
 m_\varphi^2 
= - \frac{ 4\pi m_q }{\Nc} \la \bar{\psi} \psi \ra \, .
\end{equation}
%
\section{One baryon, two baryons, and finite baryon density
(One flavor)}
\label{baryon}
From this section,
we analyze the finite density problems
using the non-Abelian bosonization.
We will first discuss the one flavor case,
and the two flavor extension will be discussed
in Sec.\ref{2flavordense}.

While our main concern in this paper is
the properties of Quarkyonic matter,
perhaps it is instructive to discuss
the properties of baryonic matter 
in light of the present framework.
Therefore we will start our discussions
from a single baryon.
Then we argue a baryon-baryon interaction,
and finally move to the Quarkyonic matter. 

The outline our discussions is the following:
The ground state baryon is constructed as a soliton
since we fully bosonize the fermion operators.
The solitonic construction turns out to be
less problematic in (1+1) dimensions than in (3+1) dimensions.
Unfortunately, in spite of the presence of the confining force,
the ground state baryon in QCD$_2$ tells us little about
the $\Nc$-quarks bound by the color fluxes.
We will interpret this result 
in light of the charge-color separation,
which is absent in (3+1) dimensions.
So we can not discuss detailed structual changes
of baryons in terms of fermionic contents or flux tubes,
in a manner applicable to the (3+1) dimensional 
cases\footnote{The fermionic description
is found in Ref. \cite{Schon:2000he}.
The energy contains the IR divergent piece
as a consequence of the confining models.
Without confinement, the single fermion properties
become well-defined due to the smallness of the
residual interactions.
An interested reader should consult with,
for instance,
papers \cite{Schon:2000he,Schnetz:2005ih} 
for the Gross-Neveu model,
papers \cite{nucl-th/9609012}
for the NJL model for baryons,
and a recent paper \cite{arXiv:1008.4029}
for the nuclear-quark matter transition.
}.
We also see that a single baryon 
accompanies a single chiral spiral.
After showing the operator relation special to (1+1) dimensions,
we will give its qualitative picture
which, to some extent, may be extended to 
the (3+1) dimensional considerations.

Next the baryon-baryon interaction is discussed.
At large $\Nc$,
the strongest force originates from the
$O(\Nc)$ quark density, leading to
the coherent meson exchange with the large 
amplitude\footnote{All other meson exchanges happen only 
as quantum processes, so amplitudes 
are suppressed by $1/\Nc$.}.
The force is purely repulsive,
and we did not find the attractive part
which, in (3+1) dimensions,
may emerge from the $\sigma$ 
exchange\footnote{If the $\sigma$ meson entirely
originates from the correlated $2\pi$ exchanges,
it should be assigned as
quantum processes in the present framework.}.
At higher density,
this force from $U(1)$ charges will remain important,
in sharp contrast to the forces from 
the non-Abelian charges like isospins which 
will eventually cancel out one another.
Thus irrespective of contents of flavors,
the relevance of the 
strong repulsive force grows as increasing density, 
eventually invalidating the baryon based descriptions.
We will enter the regime where the quark picture
is necessary to account for the bulk part of the Fermi sea.

Finally we argue the Quarkyonic matter regime.
The basic degrees of freedom are 
a quark and a quark-hole which form
the color singlet mesonic objects.
The condensations of them provide the chiral spirals.
In spite of the homogeneous distributions
of chiral densities,
quark number density is nearly uniform.
We will also see the
fundamental excitations are particles and holes,
and there is no strong motivation 
to consider baryonic excitations.
The $n$ ($n$: positive integer)
particle-hole picture
provides much more general descriptions
than a baryon and a baryon-hole excitation,
or a baryonium bound state.

\subsection{One baryon as a topological object}
\begin{figure}[b]
\begin{center}
\scalebox{1.0}[1.0] {
  \includegraphics[scale=0.7]{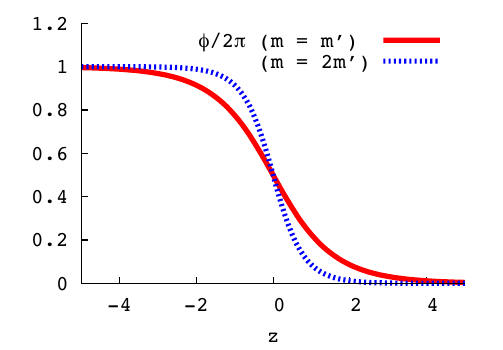} }
\scalebox{1.0}[1.0] {
  \includegraphics[scale=0.7]{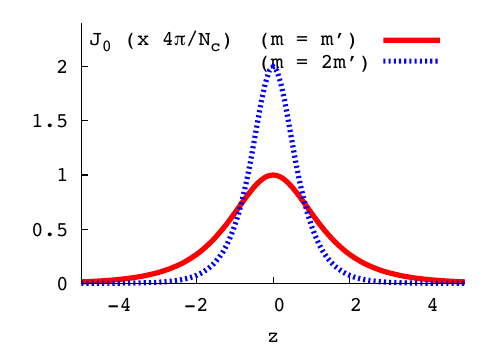} }
\end{center}
\caption{The behaviors of
$\phi$ field devided by $2\pi$ (left panel) 
and quark number current $j_0$ devided by $\Nc/4\pi$
(right panel)
as functions of $z$ with some arbitrary dimensionful unit, $m'^{-1}$.
We chose $z_0$ to be $0$.
For each plot, we took $m_\varphi = m'$ 
and $m_\varphi=2m'$
to see the mass dependence.
}
\label{baryonfig}
\end{figure}

With a boundary condition for a topological charge,
we can find a solution for
coherent configurations $\bar{U}= \rme^{\rmi \phi}$
regarding $\phi$ as $O(\Nc^0)$ quantity.
The action becomes the sine-Gordon model,
\begin{equation}
S^{\eff}[\phi]
 = \frac{\Nc}{8\pi} \int \rmd^2 x 
\big[ (\partial_\mu \phi)^2
 + 2 m_\varphi^2 (\cos{\phi}-1) \big] \, ,
\end{equation}
whose properties are well-investigated \cite{Colemanbook}.
At large $\Nc$, stationary phase approximation
is applicable due to an overall $\Nc$ factor of the action. 
Using Bogomol'nyi's trick, we have the lower bound of the energy,
\begin{align}
E^{\eff}[\phi]
 &= 
\frac{\Nc}{8\pi} \int \rmd z
\bigg( \partial_z \phi \mp 2m_\varphi \sin \frac{\phi }{2}\bigg)^2 
\pm \frac{\Nc m_\varphi}{2 \pi} 
\int^{\phi(+\infty)}_{\phi(-\infty)}
 \!\! \rmd\phi~ \sin \frac{\phi }{ 2} \nonumber \\
&\ge 
|N_B| \times \frac{2 \Nc m_\varphi}{\pi} \, , 
\label{integral}
\end{align}
where $\phi(\pm \infty)$ must be $2\pi\times$ integer.
The equality holds only when the first
bracket becomes zero.
It is satisfied only by the 
topological charge one solution,
whose configuration and energy are,
\begin{equation}
\phi(z;z_0) = 4 \tan^{-1} \rme^{-m_\varphi (z-z_0)} \, ,
~~~ E= \frac{2\Nc m_\varphi}{\pi} \, ,
\label{chiraldensity}
\end{equation}
where $z_0$ is a free parameter
to characterize a center of a single baryon.
Here $\phi$ goes to $0$ as $z \rightarrow \infty$
and to $2 \pi$ as $z \rightarrow - \infty$,
as shown in the left panel of Fig.\ref{baryonfig}.
The baryon density is localized 
around $z= z_0$ as (the right panel of Fig.\ref{baryonfig})
\begin{equation}
J_0 = -\frac{\Nc}{2\pi} \partial_z \phi
= \frac{ \Nc m_\varphi }{ 4 \pi \cosh \big( m_\varphi(z-z_0) \big)} 
\, ,
\label{baryondensity}
\end{equation}
By integrating $J_0$ with respect to $z$,
we can see its quark number is $\Nc$.

With expressions (\ref{chiraldensity}) and (\ref{baryondensity}),
the relationship between
a quark number and chiral spirals is explicit.
The scalar and pseudoscalar chiral 
densities are
(here we attach a minus sign in front of
$\Delta\equiv |\la \bar{\psi} \psi\ra_{{\rm VAC}}|$ 
since $\la \bar{\psi} \psi \ra_{{\rm VAC}}<0$),
\begin{align}
\la \bar{\psi} \psi (z) \ra_B
&=
\la \bar{\psi}_+ \psi_- \ra_B 
+ \la \bar{\psi}_- \psi_+ \ra_B 
= - \Delta \cos \phi(z;z_0) ~, 
\\
\la \bar{\psi} \rmi \gamma_5 \psi (z) \ra_B
&=
- \rmi \big(\la \bar{\psi}_+ \psi_- \ra_B
- \la \bar{\psi}_- \psi_+ \ra_B \big)
= - \Delta \sin \phi(z;z_0) \, ,
\end{align}
so we find one chiral spiral around a single baryon.
It is natural to expect more chiral spirals
when we have a larger number of baryons,
or more generically, quark number density.
We will see this explicitly later.
\begin{figure}[b]
\begin{center}
\scalebox{1.8}[1.0] {
\hspace{0.2cm}
  \includegraphics[scale=0.16]{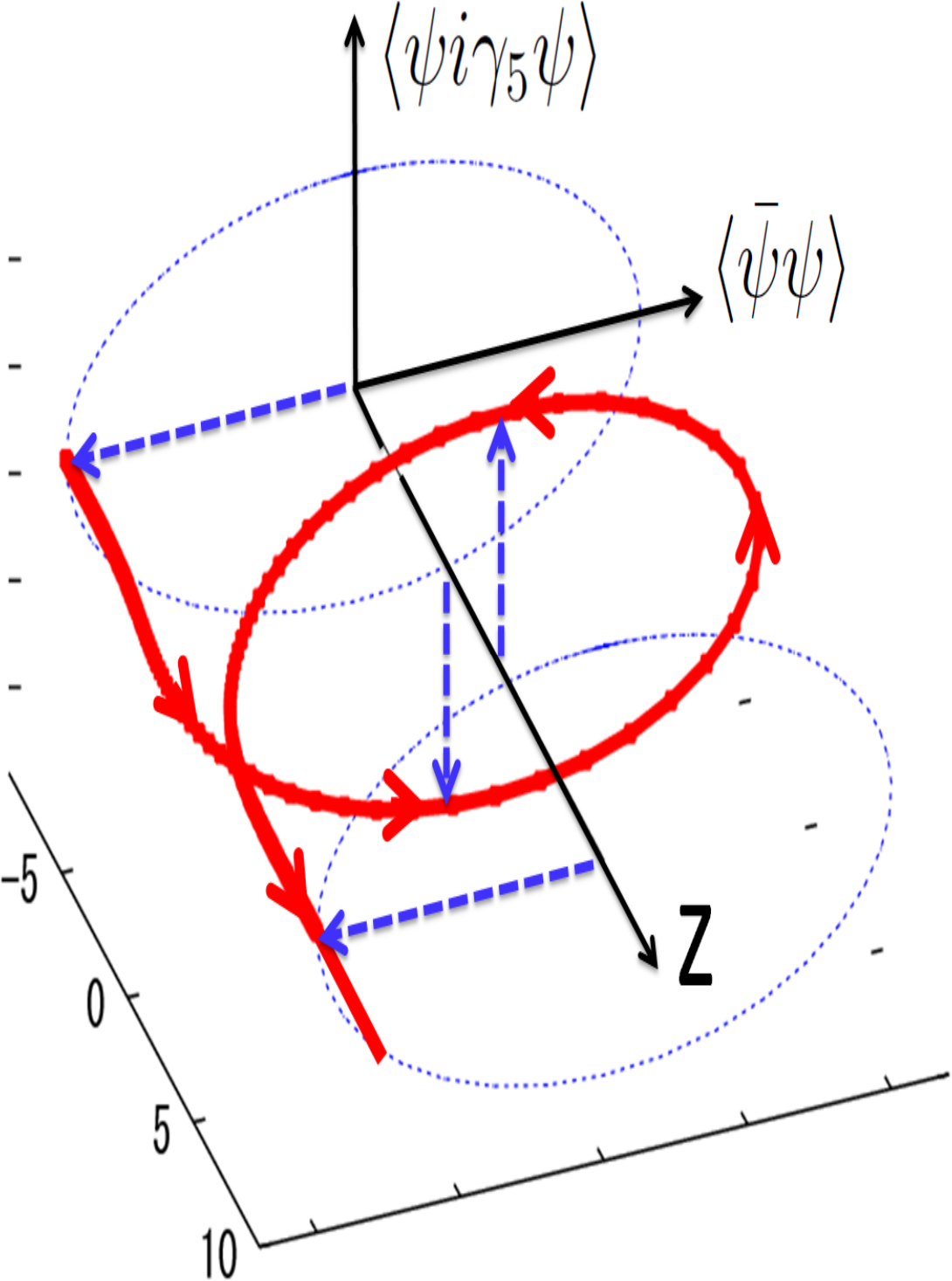} }
\scalebox{1.0}[1.0] {
\hspace{-0.2cm}
  \includegraphics[scale=0.7]{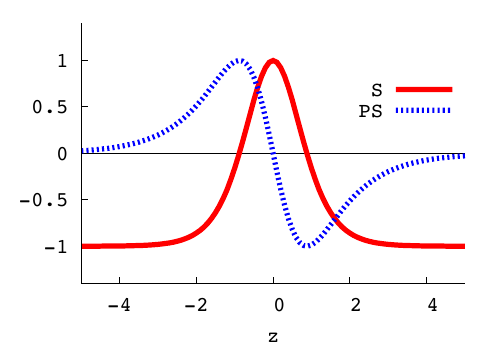} }
\end{center}
\caption{The behaviors of
scalar (S) $\la \bar{\psi}\psi \ra$ 
and pseudo scalar (PS) $\la \bar{\psi} \rmi \gamma_5 \psi \ra$
chiral condensates as functions of $z$.
The unit of condensates is $\Delta$.
Left panel: 3D plot of the chiral spiral.
Right panel: Separate plots of condensates.
}
\label{chiralfig}
\end{figure}

For future references,
perhaps it is useful to rephrase the emergence of
chiral spirals in a different way,
without emphasizing 
(1+1) dimensional operator relations.
Note that $\la \bar{\psi}_+ \psi_- \ra$ 
($\la \bar{\psi}_- \psi_+ \ra$) expresses
an average density of pairs co-moving
in the $-z$ ($+z$) direction.
Baryons make different potentials for 
pairs moving in opposite directions.
Accordingly there is a mismatch
in densities $\la \bar{\psi}_+ \psi_- \ra$ 
and $\la \bar{\psi}_- \psi_+ \ra$
near the baryon,
so that $\la \bar{\psi} \rmi \gamma_5 \psi \ra$
does not vanish.
We can repeat similar arguments
in higher dimensions, once we 
replace $\gamma_5$ with $\gamma_0 \gamma_j$,
whose eigenvalues characterize
moving directions of particles
in a similar way as $\gamma_5$ does
in (1+1) dimensions \cite{Kojo:2009ha,Son2000}. 

Finally we would like to examine
qualitative differences in
solitonic constructions of baryons in
(1+1) and (3+1) dimensions.
In the former, computations
can be closed within dynamics around the scale,
$m_\varphi \ll \lqcd$.
This is in sharp contrast to (3+1) dimensional 
cases \cite{Adkins:1983ya}:
To get stable soliton solutions in the chiral Lagragian,
we need to equate the leading order with the
next leading order 
of the derivative expansions, $(\sim \partial/\lqcd)^n$.
But once the LO and NLO become comparable,
then, in principle, all possible higher orders 
become relevant \cite{Weinbergbook}.
The effects of higher orders are needed to 
be computed or be replaced with quark degrees of freedom.

Another important difference is that the former
can be constructed from $U(1)$ bosons only, 
instead of flavored pions in (3+1) dimensional solitons.
It is natural to have the 
coherent configurations, $O(\Nc)$ 
amplitude of $U(1)$ bosons, since quark number density 
is $O(\Nc)$ inside of the baryon.

Taking into account these differences between
(1+1) and (3+1) dimensions,
the former case has less problems to apply the solitonic picture to 
the ground state baryon. 
Presumably this simplicity comes from the fact that 
we see remnants of the charge-color separation.
Such a separation is broken only via a term 
proportional to a small current quark mass,
and the dynamics of colors only plays a very indirect role. 
Thus we expect that
baryons discussed here have little to do with
the picture of $\Nc$-quarks bound by the color fluxes.
This means that
our approach can not answer to several interesting questions
such as how quark wavefunctions or color flux tubes
inside of baryons change as increasing density.

\subsection{Two baryons and interactions}

As we already saw in the one baryon sector,
Bogomol'nyi's bound is satisfied
only by a single baryon solution.
It means that
the energy of two baryons is always larger than
twice of single baryon energy,
\begin{equation}
E_{B=2} > 2 E_{B=1} \,.
\end{equation}
The additional energetic cost comes from
the repulsive interaction between two baryons.
Its asymptotic form is
($R$ is the distance between two baryons)
\begin{equation}
V(R) \sim \Nc m_\varphi \rme^{-m_\varphi R} \, .
\end{equation}
When $R\sim m_\varphi^{-1}$,
the strength of repulsion
becomes the same order as the single baryon mass.

How generic is this sort of the strong repulsive
force in general dimensions?
In QCD$_4$, we know that the $\omega$ meson exchanges
in nuclear forces are strong and repulsive.
One interpretation of the strength is that
it comes from the quark number of $O(\Nc)$
which produces a large number of $U(1)$ mesons
of $O(\Nc)$.
The repulsive feature presumably comes from
an additive property of $U(1)$ charges,
in sharp contrast to other non-Abelian charges
such as isospins
which stays at $O(\Nc^0)$ for the ground state baryons.

We expect that
this viewpoint be increasingly important
at higher quark density.
While non-Abelian charge densities such as colors or
isospins cancel one another and remain at $O(1)$,
the enhancement of the quark number density 
is not tempered at large density.
Then strong repulsive
interactions become increasingly important,
and then completely change baryon structures.
Eventually computations based on the quasi-particle
picture of baryons break down,
implying that the quarks become
reasonable effective degrees of freedom
to describe most part of the Fermi sea.

\subsection{Finite density (one flavor)}
\label{1flavordense}
Since two baryon interactions are repulsive,
baryonic matter starts to appear from
the critical quark chemical potential,
$\mu_c \equiv M_B/\Nc = 2m_\varphi/\pi \sim 0.7 m_\varphi$.
In the following, the canonical approach
is used.
Since we have a large number of quarks,
the leading contributions 
can be computed by studying coherent field
configurations.
With the coherent field expression $U\sim e^{i\phi}$,
a total quark number of the system per period $L$ 
is
\begin{equation}
\hspace{-0.5cm}
N_q = - \int^L_{0} \rmd z J_0[\phi] 
= - \frac{\Nc}{2\pi} \int_{0}^{L} \rmd z~ \partial_z \phi
= - \frac{\Nc}{2\pi} \big[ \phi(L) - \phi(0)\big]
= \Nc N_B \, .
\label{baryoncharge}
\end{equation}
Below we take $\phi(L) - \phi(0) = - 2\pi N_B$
so that $n_B= N_B/L$ represents a baryon 
density\footnote{ 
We require quantum fluctuations $\varphi$ 
to be normalizable modes, 
so that the expression (\ref{baryoncharge}) saturates
the average baryon density.}.
We will use $p_f = \pi n_B$ in the following.

With this constraint of a finite quark number,
we first determine a static background
in the large $\Nc$ stationary phase approximation.
The Hamiltonian for static coherent fields 
is\footnote{We assume that $\tilde{\phi}$ does not 
contain any constant,
since it can be always absorbed by
shifting $2 p_f z \rightarrow 2 p_f (z-z_0)$.}
\begin{align}
\frac{ E[\phi] }{L}
&= \frac{\Nc}{4\pi L} \int \rmd z 
\bigg( \frac{1}{2} (\partial_z \phi)^2
 - m_\varphi^2 (\cos{\phi}-1) 
\bigg) \nonumber \\
&=
\frac{\Nc}{2\pi}( 2m_\varphi^2 + p_f^2)
+ 
\frac{\Nc}{8\pi L} \int \rmd z 
\bigg\{ (\partial_z \tilde{\phi} )^2
 - 2 m_\varphi^2 \cos (\tilde{\phi} - 2 p_f z )
\bigg\} \, ,
\label{clenergy}
\end{align}
where we have decomposed $\phi(z)$ field into
$\phi_0 (z) \equiv - 2 p_f z$ 
and $\tilde{\phi}(z)$.
So $\tilde{\phi}$ satisfies a periodic
boundary condition, 
$\tilde{\phi}(0)= \tilde{\phi}(L)$,
forming the manifold $S_1$.
To find a solution for coherent configurations,
we have to solve the following equation,
\begin{equation}
\partial_z^2 \tilde{\phi} 
- m_\varphi^2 \sin(\tilde{\phi}- 2 p_f z)= 0 \, .
\label{EOM}
\end{equation}
The equation has been investigated extensively.
It is known that solitonic configurations are 
general solutions which interpolate
well-separated baryonic soliton configurations 
at low density
and uniform quark number distributions at high density
(For intensive discussions, see \cite{Schnetz:2005ih}).

We will not repeat detailed analyses here 
but just quote some results
about high density
to assure our qualitative discussions in Sec.\ref{general}.

In chiral limit, $\tilde{\phi}(z)=0$ is the solution
which means that the baryon density is uniform.
The colored sector and the Lagrangian 
for quantum fluctuations are exactly the same as 
the vacuum. 
The chiral condensate behaves as
$\la \bar{\psi}_\mp \psi_\pm \ra
= \Delta \rme^{\pm \rmi \phi} = \Delta \rme^{\pm 2 \rmi p_f z}$,
and form the chiral spirals.
The coefficient $\Delta$ equals to
the vacuum value\footnote{
It might seem strange that we get chiral spirals
without any energetic minimization with respect to
the chiral condensate. 
Actually, the modulus of chiral condensates
is served from the color sector,
independently from properties of the $U(1)$ sector.
If one hopes to see energetic differences
with or without chiral condensates,
one must derive an expression of 
the energy in the color sector.
}. 
The classical energy density is
\begin{equation}
\frac{ E[\tilde{\phi}=0] }{L} 
= \frac{\Nc}{2\pi} p_f^2 \, .
\end{equation}
The density contributions to the energy 
density is just those of free fermions.

Next let us consider the case of massive fermions
in asymptotically high density.
If we rewrite Eq.(\ref{EOM}) by 
scaling a variable as $z'=2 p_f z$,
\begin{equation}
\partial_z'^2 \tilde{\phi} 
+ \bigg( \frac{m_\varphi }{ 2 p_f} \bigg)^2
\sin(z' - \tilde{\phi})= 0 \, ,
\label{EOM2}
\end{equation}
so $\tilde{\phi} \sim (m_\varphi/p_f)^2$,
and we can organize expansions. 
An asymptotic behavior,
\begin{equation}
\tilde{\phi}^{(1)} 
\simeq 
\bigg( \frac{m_\varphi }{ 2p_f} \bigg)^2 \sin(2p_f z) \, ,
\label{expression}
\end{equation}
satisfies Eq.(\ref{EOM}) up to $O(m_\varphi^2/p_f^2)$.
This expression illustrates that
roles of mass become less relevant as 
density increases.

\begin{figure}[t]
\begin{center}
Quark number density $j_0$
\end{center}

\vspace{0.15cm}
\hspace{2.1cm} {\small $p_f/\mu_c=0.5$}
\hspace{1.8cm} {\small $p_f/\mu_c=1.0$}
\hspace{1.82cm} {\small $p_f/\mu_c=1.5$}
\vspace{-0.35cm}

\begin{center}
\scalebox{1.0}[1.0] {
\hspace{0.0cm}
  \includegraphics[scale=0.8]{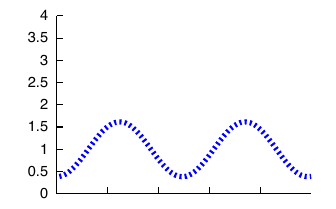} }
\scalebox{1.0}[1.0] {
\hspace{-0.5cm}
  \includegraphics[scale=0.8]{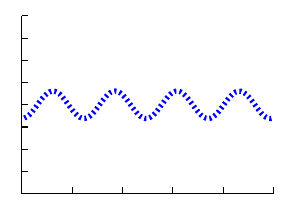} }
\scalebox{1.0}[1.0] {
\hspace{-0.55cm}
  \includegraphics[scale=0.8]{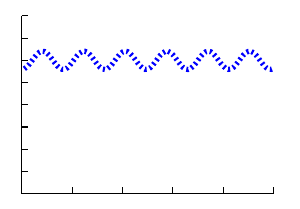} }
\end{center}

\vspace{0.2cm}
\begin{center}
Chiral condensate $\la \bar{\psi} \psi \ra$
\end{center}
\vspace{0.2cm}

\begin{center}
\scalebox{1.0}[1.0] {
\hspace{0.0cm}
  \includegraphics[scale=0.8]{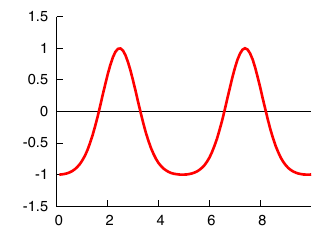} }
\scalebox{1.0}[1.0] {
\hspace{-0.5cm}
  \includegraphics[scale=0.8]{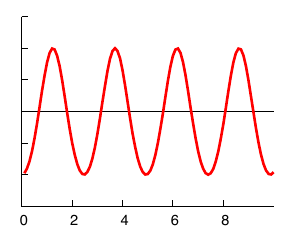} }
\scalebox{1.0}[1.0] {
\hspace{-0.55cm}
  \includegraphics[scale=0.8]{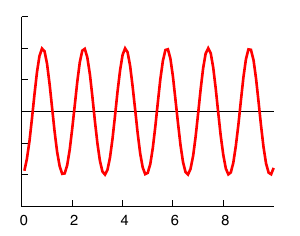} }
\end{center}

\begin{center}
\hspace{0.5cm}
$z$
\end{center}

\caption{The behaviors of
quark number density (upper panels)
and chiral scalar density (low panels)
as functions of $z$ with different quark densities
$p_f/\mu_c=0.5,\ 1.0,\ 1.5$.
(The units of $z,\ j_0,\ \la\bar{q}q\ra$
are $m_\varphi^{-1}$, $\Nc\mu_c/2\pi$,
$\Delta$, respectively.)
}
\label{densefig}
\end{figure}

With the expression (\ref{expression}),
we can compute a number of quantities of interest
(Summary of distributions can be found in Fig.\ref{densefig}.).
The oscillation of the chiral condensate is slightly 
deformed as
\begin{equation}
\la \bar{\psi} \psi \ra
\simeq 
\Delta \cos( \phi_0 + \tilde{\phi}^{(1)} )
=
\Delta \cos\bigg( 2p_f z 
 - \frac{m^2}{4 p_f^2} \sin(2 p_f z)\bigg) \, ,
\label{densitycondensate}
\end{equation}
and $\la \bar{\psi} \rmi \gamma_5 \psi \ra$ 
also oscillates in the similar way.
The baryon density also acquires modulations,
\begin{equation}
 \la J_0 \ra 
\simeq 
- \frac{\Nc}{2\pi}\partial_z [ \phi_0 + \tilde{\phi}^{(1)} ]
= \frac{\Nc}{2\pi} 
\times 
2 p_f \bigg\{1 - \frac{m_\varphi^2}{4 p_f^2} \cos(2 p_f z) 
\bigg\} \, .
\label{densitybaryon}
\end{equation}
In average, this solution does not contribute
to the total baryon charge,
since it just fluctuates around zero.
At high density, 
prominent structures in quark number density 
is eventually buried in the average quark number background.
The average energy density for 
$\tilde{\phi} = \tilde{\phi}^{(1)}$ is
\begin{equation}
 \frac{E[\tilde{\phi}^{(1)}]}{L}
= 
\frac{\Nc p_f^2}{2\pi} 
\bigg\{ 1
+ 2 \bigg( \frac{ m_\varphi }{ p_f } \bigg)^2 
- \frac{1}{32}
 \bigg( \frac{ m_\varphi }{ p_f } \bigg)^4 
+ O(m_\varphi^6/p_f^6)
 \bigg\} \, ,
\label{densityenergy}
\end{equation}
where the second term comes from the vacuum constant,
while the third term expresses 
corrections from the density wave modulations.

\begin{figure}[tb]
\begin{center}
\scalebox{0.5}[0.5] {
\hspace{-1.0cm}
  \includegraphics[scale=0.45]{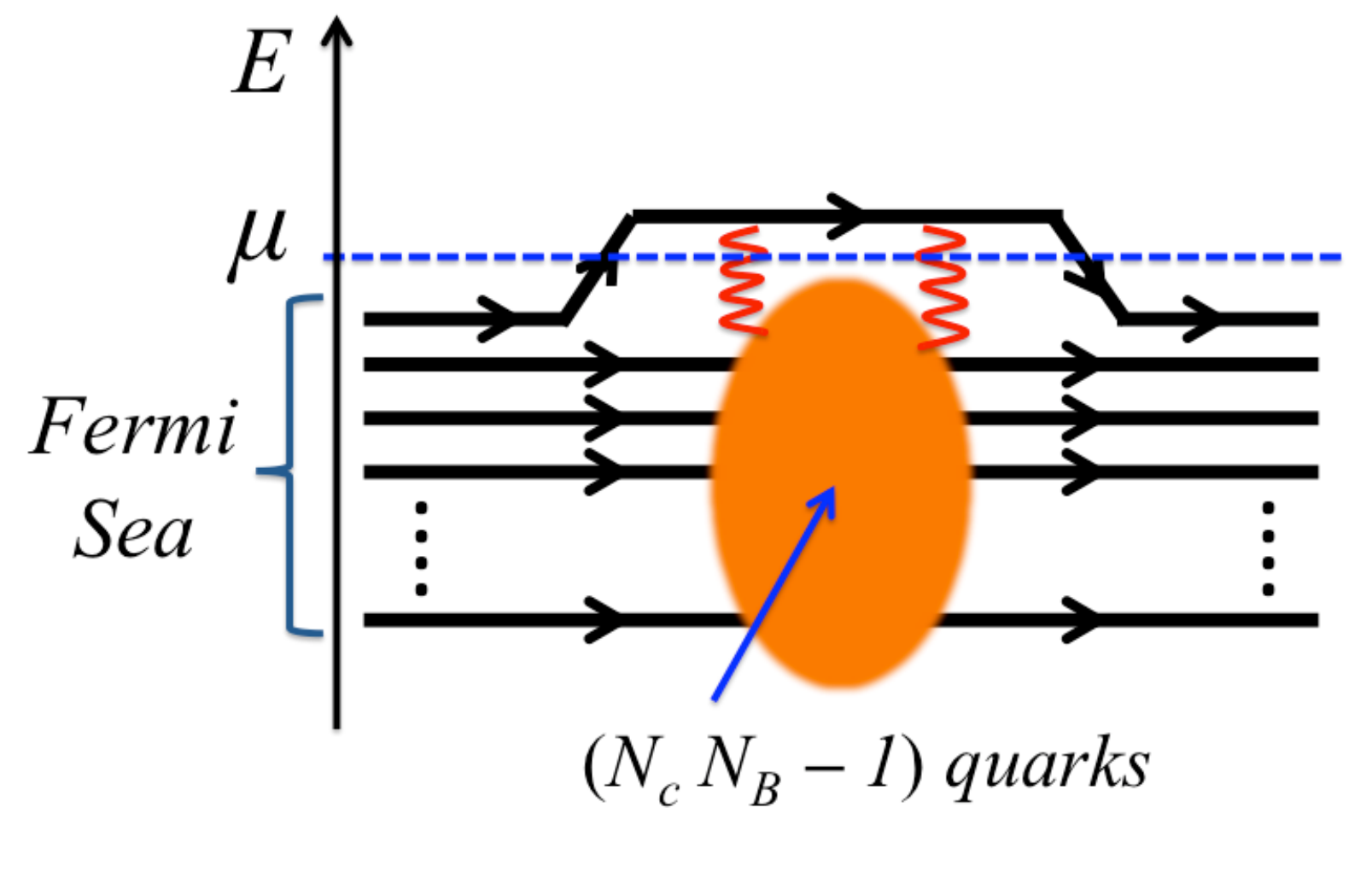} }
\hspace{1.0cm}
\scalebox{0.5}[0.5] {
  \includegraphics[scale=0.45]{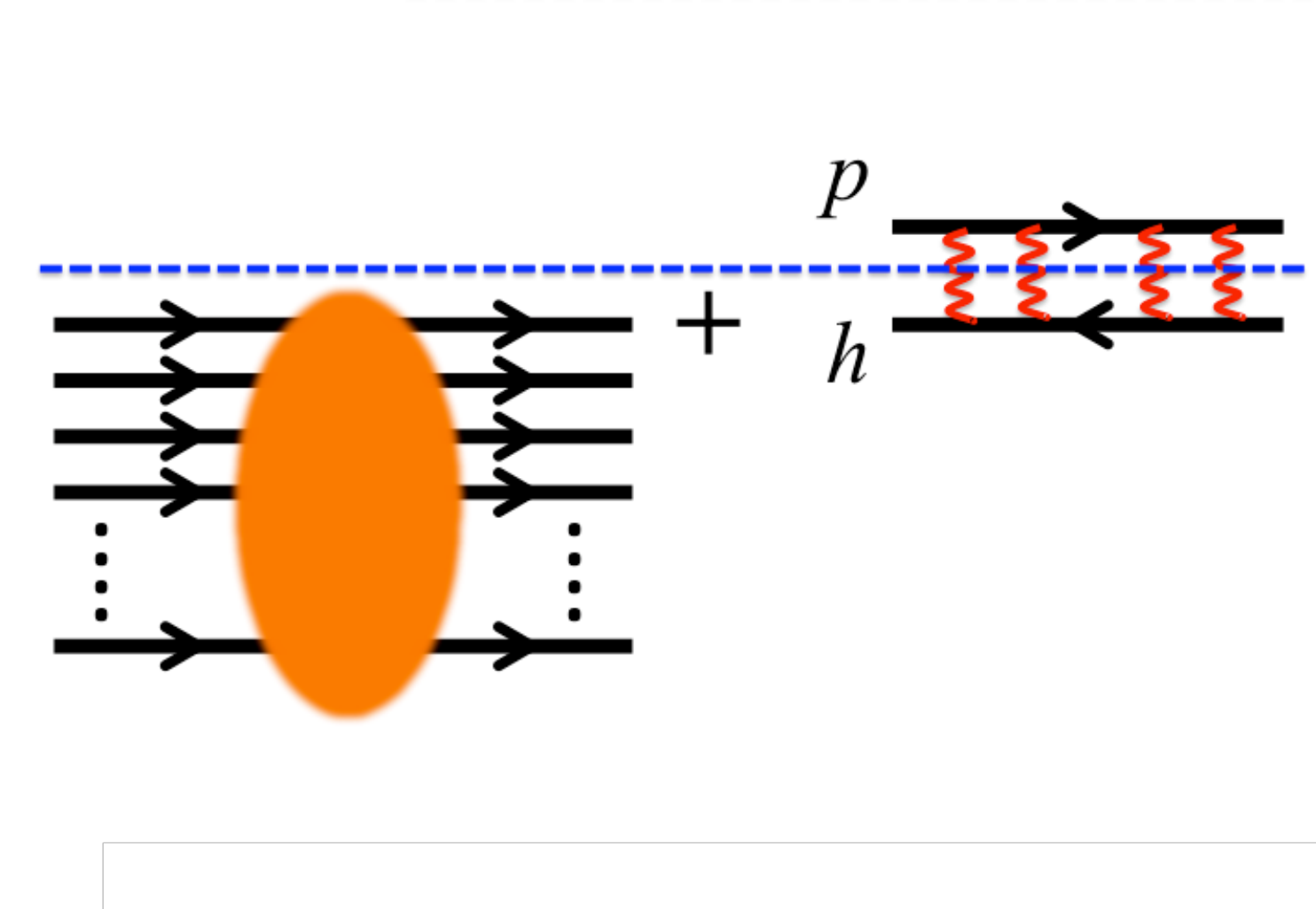} }
\end{center}
\caption{A schematic picture for a particle-hole exciation.
The case with total baryon number, $N_B$, is illustrated.
(Left) The interactions between one quark excitation and
$(\Nc N_B - 1)$ quarks.
(Right) The equivalent diagram in the particle-hole picture.
}
\label{ph}
\end{figure}

An easy way to interpret
expressions (\ref{densitycondensate}),
(\ref{densitybaryon}), and (\ref{densityenergy}),
in the fermionic language
is to employ a particle-hole picture
in the quark Fermi sea (Fig.\ref{ph}).
When baryons largely overlap,
quark number distribution is almost uniform.
Deviations from such distributions 
can be regarded as corrections caused by 
particle-hole degrees of freedom.
Indeed, spatial modulations have a period $1/2p_f$,
reflecting that condensations are driven by
co-moving particle-hole pairs near the Fermi surface.
Note also that 
the shape of each peak in the baryon density (\ref{densitybaryon})
is very different from that of the single baryon solution.

Based on the above arguments,
it seems better to regard
baryonic configurations just as
quark number localizations and defects,
made of $n$ particle-hole pairs,
where $n$ is a positive integer.
They contains a wider class of excitations than
baryon and baryon-hole excitations,
because $n$ need not to be 
quantized to a particular number, $\Nc$.
Therefore we have no strong motivation
to focus on baryon excitations in usual sense,
in which the color-singletness is maintained 
within a single baryon or a single baryon-hole, 
by definition.

One might think a possibility of a baryon
and a baryon-hole bound state and
its condensation.
But it should be described not only by
the $\Nc$ particle-hole pairs,
but also by $\Nc-1$, $\Nc-2$, ...
particle-hole pairs because of the
partial annihilations of particle-holes.
Again we have no good reason to start with
baryons and baryon-holes for the
considerations of the condensation phenomena.

A more nontrivial possibility is the dibaryon type
condensation such as baryon superfluidity.
But in QCD$_2$ with one flavor,
we observed only the repulsive force at large $\Nc$,
so such phenomena can occur only after the subleading
effects of $1/\Nc$ are included.
To explore this possibility for finite $\Nc$, 
we need quantitative arguments.
We leave it for future studies.

With combining all these pictures together,
at least in the large $\Nc$ limit,
our system is naturally viewed as a system
with the Fermi sea of weakly coupled quarks
plus mesonic excitations of particle-hole type.
These arguments are consistent with
discussions in Sec.\ref{general}
which are based on the fermionic language.

\section{Two flavors}
\label{2flavordense}
In this section,
we extend analyses to two flavor cases with $u$ and $d$ quarks,
keeping the vector symmetry $SU(2)_V$ by taking
equal current quark masses $m_q = m_u = m_d$.
While most of treatments are the same as before,
it is interesting to see
the possibility of the chiral spirals
including the flavor rotations
such as $\la \bar{\psi} \rmi \gamma_5 \tau_3 \psi\ra$.
This issue may have phenomenological relevance
to stars via the spontaneous generation of 
magnetic fields of the QCD scale.

According to Vafa-Witten theorem \cite{Vafa:1983tf}
at zero density,
the vector-like theories
with current quark masses can not 
have the flavor symmetry breaking.
However, several conditions used in that proof
can not be applied in the presence of the chemical potential, 
so this problem is interesting in its own right.
In the following, we will not turn on electromagnetic
interactions, but just see what happens in the QCD sector.
(In case of NJL$_2$ studies in the presence of
 isospin chemical potential,
see Ref. \cite{Ebert:2011rg}.)

We will see that, in the present model,
the chiral spirals mainly 
occur between $\la \bar{\psi} \psi \ra$
and $\la \bar{\psi} \rmi \gamma_5 \psi \ra$.
The rotations including 
$\la \bar{\psi} \rmi \gamma_5 \tau_a \psi \ra$
can also occur, 
but it is order of 
$m_\varphi^2/\mu^2$ compared to $U(1)$ channel,
so will eventually disappear at high density.
This can be interpreted as
cancellations of non-Abelian charges at high density.

\subsection{Vacuum}
When we consider the multi-flavors,
we have only to add slight modifications
to previous treatments.
The coefficients of $U(1)$ and color currents change as,
\begin{equation}
J_- (x) 
= \rmi \frac{\Nc \Nf }{ 4\pi} U \partial_- U^\dag \, ,
~~~
J^A_- (x) 
= \rmi \frac{\Nf }{ 2\pi} \tr[ h \partial_- h^\dag t_A] \, ,
\end{equation}
and flavor currents are defined as
\begin{equation}
J^f_- (x) 
 = \rmi \frac{\Nc }{ 2\pi} 
   \tr\big[ g \partial_- g^\dag \frac{\tau_f }{ 2}\big] \, ,
~~~
J^f_+ (x) 
 = \rmi \frac{\Nc }{ 2\pi} 
   \tr\big[ g^\dag \partial_+ g \frac{\tau_f }{ 2}\big] \, ,
\end{equation}
where the $g$ is a matrix field and $\tau_f$ is a flavor matrix.
with normalization $\tr[\tau_f \tau_{f'}]= 2 \delta_{ff'}$.
The operators for the chiral density are
\begin{equation}
\bar{\psi}^{ai}_+ \psi^{bj}_- 
= c(M)~ (h^{ab})_M (U)_M (g^{ij})_M \, ,
\end{equation}
where $i,j, \dots$ are flavor indices 
for the fundamental representation.
The action for the fermion kinetic term is
\begin{equation}
 \int \rmd^2x \,
 \bar{\psi}(x) \rmi \Slash{\partial} \psi (x) \ 
\longrightarrow \ 
 S^{ {\rm U(1)} }_{k=\Nc \Nf} [U] 
+ S_{k=\Nc}^{{\rm WZW}}[g] + S_{k=\Nf}^{{\rm WZW}}[h] \, .
\end{equation}
The treatments of the colored interactions 
are the same as before,
and in chiral limit,
there is the charge-flavor-color separation.

Below we will consider $\Nf=2$, massive quarks.
With current quark masses, we have the following terms
after applying a probe approximation
and using $\tr[g]=\tr[g^\dag]$,
\begin{align}
{\mathcal L}_m
&= - m_q (\bar{\psi}_+ \psi_- + \bar{\psi}_- \psi_+ )
\nonumber \\
&\rightarrow  \frac{\Nc m_\varphi^2 }{ 8\pi} \,
\tr[g](U + U^\dag )
=
\frac{\Nc \Nf m_\varphi^2 }{ 4\pi} \cos\Pi \cos \phi \, ,
\label{flavormass}
\end{align}
where we parameterized $U=e^{i\phi}$, and
for the matrix field,
\begin{equation}
g = \rme^{ \rmi \, \Pi_f \tau_f}
= \cos \Pi + \rmi \, \hat{\Pi}_f \tau_f \sin \Pi
\hspace{1cm}
 (\Pi^2= \Pi_f^2,\ \ \hat{\Pi}_f = \Pi_f/\Pi) \, .
\end{equation}
This simple expression holds for $\Nf=2$,
since $\{\tau_f,\tau_{f'} \}=2\delta_{ff'}$.

From the form of Eq.(\ref{flavormass}),
we can see that the trivial vacuum is found by choosing
$(\phi,\Pi) = (0,0)$ and $(\pi,\pi)$ (modulo $2\pi$),
which maximizes the absolute value
of the chiral condensate.
On the other hand,
the flavor or parity violating chiral condensates
are
\begin{align}
\bar{\psi}\, \rmi \gamma^5 \psi 
\ &\sim \ 
\tr [g] \left( \rme^{\rmi \phi} - \rme^{-\rmi\phi} \right)
\ \sim \ \cos\Pi \sin \phi \,,
\label{flavorcon1}
\\
\bar{\psi} \tau_f \psi 
\ &\sim \ 
\left( \tr [g \tau_f]\, \rme^{\rmi \phi} 
+ \tr [g^\dag \tau_f]\, \rme^{-\rmi \phi} \right)
\ \sim \ 
\hat{\Pi}_f \sin\Pi \sin \phi \,,
\label{flavorcon2}
\\
\bar{\psi}\, \rmi \gamma^5 \tau_f \psi 
\ &\sim \ 
\left( \tr [g \tau_f] \, \rme^{ \rmi\phi} - \tr [g^\dag \tau_f] \,
  \rme^{-\rmi \phi} \right)
\ \sim \
\hat{\Pi}_f \sin\Pi \cos \phi \,,
\label{flavorcon3}
\end{align}
becomes zero.
The above expressions indicate that
these condensates compete with
the usual chiral condensate.
The usual chiral condensate wins 
because of the mass term.

The action for quantum excitations 
can be obtained by expanding fields around
this vacuum.
Taking $U = \rme^{ \rmi (4\pi/\Nc\Nf)^{1/2} \varphi}$
and $g = \rme^{ \rmi (2\pi/\Nc)^{1/2} \pi_f \tau_f}$,
we have
\begin{equation}
\!\!\!\!
S^{{\rm eff}}[\varphi,\pi_f]
= \int \rmd^2 x~
 \frac{1}{2} 
\left( (\partial_\mu \varphi)^2 
+ (\partial_\mu \pi_f)^2
- m_\varphi^2 
(\varphi^2 + \pi_f^2 )
\right) + O(1/\Nc^{1/2}) \, .
\end{equation}
The masses of bosons in $U(1)$ and $SU(2)$
are degenerate as a consequence
of large $\Nc$.
At finite $\Nc$, there are higher order 
interaction terms in $SU(2)$ sector,
and they destroys such a degeneracy.

\subsection{Finite density}
Calculations of finite density case
can be done as before.
We again consider the canonical approach,
and take the average isospin density zero.
Then we have a periodic boundary condition,
$\Pi_f(0)=\Pi_f(L)$.
The static energy functional for 
coherent field configurations
is
\begin{equation}
E[\tilde{\phi},\Pi_f] 
= \Nc \Nf \frac{\pi N_B^2}{2L}
+ E_{kin} + E_m \,,
\end{equation}
where (below we will explicitly substitute $\Nf=2$)
\begin{equation}
E_{kin}[\tilde{\phi}, \Pi_f]
= 
\frac{\Nc \Nf}{8\pi} \int \rmd z 
\left\{ (\partial_z \tilde{\phi})^2 
    + (\partial_z \Pi)^2 
    + (\partial_z \hat{\Pi}_f)^2 \sin^2 \Pi \right\} \,.
\label{clenergyflavor}
\end{equation}

In chiral limit, it is clear that
$E_{kin}$ can be minimized 
by simply taking $\tilde{\phi}=\Pi=0$.  
Thus $\phi=2p_f z$, so
chiral spirals appear only in the $U(1)$ sector.
This reflects that
an external source (a quark number constraint)
is put only in the $U(1)$ sector.

Now let us see what happens if we include
mass terms without explicitly breaking $SU(2)_V$.
The energy density from the mass term is
\begin{align}
E_m[\tilde{\phi},\Pi_f]
&= \frac{\Nc \Nf m_\varphi^2 }{ 8\pi} 
\int \rmd z \cos(2p_f z + \tilde{\phi} ) \cos \Pi
\nonumber \\
&= \frac{\Nc \Nf m_\varphi^2 }{ 16\pi} 
\int \rmd z \big( \cos(2p_f z + \tilde{\phi} + \Pi) 
+ \cos(2p_f z + \tilde{\phi} - \Pi) \big) \,.
\end{align}
We will assume that $\tilde{\phi}$ and $\Pi$ 
do not contain any constant since
they can be absorbed by shifting coordinate $z$.
And it is more enlightening to rewrite kinetic terms,
\begin{equation}
 E_{kin}[\tilde{\phi}, \Pi_f]
= \!
\frac{\Nc \Nf}{16\pi} \! \int \! \rmd z 
\left\{ \left(\partial_z (\tilde{\phi}+ \Pi) \right)^2 
+ \left(\partial_z (\tilde{\phi}- \Pi) \right) ^2 
+ (\partial_z \hat{\Pi}_f)^2 \sin^2 \Pi \right\} \,.
\end{equation}
From this form, $\partial_z \hat{\Pi}_f=0$ 
reduces the energy,
so we choose $\Pi=\Pi_3$ everywhere.
Then we have two decoupled sine-Gordon 
models with finite density,
whose variables are $\tilde{\phi}\pm \Pi_3$.
We have only to borrow results
of 1-flavor sine-Gordon model at finite density.

As in 1-flavor case,
amplitudes of $\tilde{\phi}$ and $\Pi_3$ 
are $\sim m_\varphi^2/\mu^2$ at high density.
From expressions in Eqs.(\ref{flavormass}) 
and (\ref{flavorcon1})-(\ref{flavorcon3}), 
we found the following asymptotic
behavior of condensates,
\begin{align}
\la \bar{\psi} \psi \ra,\   
 \la \bar{\psi}\, \rmi\gamma^5 \psi \ra \ 
\ &\propto \ 
\ \Delta \cos\Pi 
\ \sim \ 
\Delta \times (1 + O(m_\varphi/\mu)^2 ) \,, 
\\
\la \bar{\psi} \tau_f \psi \ra,\ 
\la \bar{\psi}\, \rmi \gamma^5\tau_f \psi \ra 
\ &\propto \ \  
\Delta \sin \Pi
\ \sim \ 
\Delta \times O(m_\varphi/\mu)^2 \,,
\end{align}
so the chiral spirals mainly occur
in the $U(1)$ sector, 
while spirals with flavor breakings
eventually disappear at high density.
On the other hand, in relatively low density,
we have solitonic lattices
with mixture of a quark number and isospins.

\section{Terms disturbing chiral spirals
\label{Impurities}}
In the preceding sections,
we saw that the chiral symmetry is broken
by chiral spirals
when we have mechanisms to keep
its modulus nonzero.
We also observed that 
an explicit chiral symmetry breaking, i.e.,
a mass term, disturbs the chiral spirals,
and such effects eventually disappear
at high density.
The purpose of this section is 
to generalize the above observation in 
such a way that we can 
classify several chiral effective models.

First let us rephrase
the role of a mass in QCD$_2$
in the fermionic language,
\begin{equation}
-m_q \int \rmd z~ ( \bar{\psi}_- \psi_+ + \bar{\psi}_+ \psi_- ) \,.
\end{equation}
Since we have the Fermi sea, it is more natural
to measure momenta and energies from the Fermi surface.
Replacing $\psi_\pm = \psi'_\pm \rme^{\pm \rmi \mu z}$,
we have
\begin{equation}
-m_q \int \rmd z~ ( \bar{\psi}'_- \psi'_+ e^{2 \rmi \mu z}
+ \bar{\psi}'_+ \psi'_- e^{- 2\rmi \mu z}) \,.
\label{massterm}
\end{equation}
Suppose that the ground state gives 
$\la \bar{\psi}'_\mp \psi'_\pm \ra \simeq {\rm const}$.
This corresponds to chiral spiral solutions
or uniform baryon distributions.
For such a state, oscillating factors $e^{\pm 2i\mu z}$ 
wash out energetic contributions from Eq.(\ref{massterm}).
This is what we actually observed in 
Eqs.(\ref{densitybaryon}) and (\ref{densityenergy}).
If $m_q$ is comparable to $\mu$,
the above state is not the ground state
and we should search solitonic configurations.
Thus the mass term disturbs
the formation of the chiral spirals.

These arguments are useful to interpret
results of other chiral models.
For instance, 4-Fermi interaction terms in 
the chiral Gross-Neveu model (NJL$_2$) 
with the continuous chiral symmetry are
\begin{equation}
H_{int} = G \int \rmd z~ \big( (\bar{\psi} \psi)^2 
+ (\bar{\psi}\, \rmi \gamma_5 \psi)^2 \big)
= 4 G \int \rmd z~ 
\left( (\bar{\psi}_+ \psi_-)(\bar{\psi}_- \psi_+) \right) \,.
\end{equation}
This form is unchanged when we rewrite fermion fields as
$\psi_\pm = \psi'_\pm \rme^{\pm \rmi \mu z}$
since oscillating factors cancel.
Therefore the model at finite density
takes the same form as that in vacuum, except
the constant energy shift associated with changes of 
the fermion fields.
The chiral spirals or uniform quark number distributions
start to appear for an infinitesimal chemical potential
-- there is no phase transition with increasing
density.

On the other hand,
the discrete Gross-Neveu model takes the following 
interaction term,
\begin{align}
H_{int} &= G \int \rmd z\, (\bar{\psi} \psi)^2 \nonumber \\
&= G \int \rmd z 
\left( 2(\bar{\psi}_+ \psi_-)(\bar{\psi}_- \psi_+) 
+ (\bar{\psi}_+ \psi_-)^2 + (\bar{\psi}_- \psi_+)^2 \right) \nonumber \\
&= G \int \rmd z  \left( 2(\bar{\psi}'_+ \psi'_-)(\bar{\psi}'_- \psi'_+) 
+ (\bar{\psi}'_+ \psi'_-)^2 \rme^{-4 \rmi \mu z} 
+ (\bar{\psi}'_- \psi'_+)^2 \rme^{4 \rmi \mu z} \right) \,.
\end{align}
The second and third terms have oscillating factors,
and disturb the formation of the chiral spirals.
Thus chiral spirals and baryons can appear
only after $\mu$ exceeds some critical value
-- there is a phase transition in contrast to
the continuous model.
At larger density,
oscillating terms become less relevant,
and the discrete model approaches to the continuous one.
Thus at high density,
this model can be regarded as NJL$_2$.

These aspects are sometimes not manifest
in the mean field energy functionals.
Let us consider the discrete Gross-Neveu model.
Typically one takes ansatz such as
$\la \bar{\psi}\psi \ra \sim 2\Delta \cos(qz)
\equiv S$. 
This can be interpreted as
\begin{equation}
( \bar{\psi} \psi )^2
= \left( ( \bar{\psi}_- \psi_+ 
- \Delta \rme^{ \rmi qz} )
+ 
( \bar{\psi}_+ \psi_- 
- \Delta \rme^{-\rmi qz} )
+ S
\right)^2 \, .
\end{equation}
But the confusing point is that
the final expression
of the mean-field energy functional is characterized
by $\la \bar{\psi}\psi \ra$ only,
and roles of 
$\la \bar{\psi}\, \rmi \gamma^5 \psi \ra$ condensate
and chiral spiral structures are hidden. 
To see it, we have to explicitly calculate
the condensate, or to construct
the effective potential
by inserting an infinitesimal external field.

Finally let us apply these arguments
to the chiral spirals or crystal structures
in higher dimensional systems.
Consider states near the region 
($p_z \sim \pm \mu,\ \vec{p}_T \sim \vec{0}_T$).
By redefining fields
$\Psi \rightarrow \rme^{ \rmi \mu z \gamma^0 \gamma^z} \Psi
= \rme^{\pm \rmi \mu z} \Psi_\pm$ 
($\Psi$ is a fermion field in higher dimensions),
the fermionic kinetic term with a chemical potential 
becomes
\begin{equation}
{\mathcal L}^{kin}
= \bar{\Psi}_+ \rmi \partial_- \Psi_+
+ \bar{\Psi}_- \rmi \partial_+ \Psi_-
+ \bar{\Psi}_+ \rmi \Slash{\partial}_T \Psi_- \rme^{- 2\rmi \mu z}
+ \bar{\Psi}_- \rmi \Slash{\partial}_T \Psi_+ \rme^{2 \rmi\mu z} \,,
\end{equation}
so transverse kintetic terms disturb
the chiral spiral rotations between
$\la \bar{\Psi} \Psi \ra$ 
and $\la \bar{\Psi} \gamma^0 \gamma^z \Psi \ra$.
This field redefinition is useful
as far as we consider only modes with
$p_T \ll 2\mu$.
It means that chiral spiral condensations
along $z$ direction 
can be made of particle-hole in the limited domain
of $p_T$.

Besides the aforementioned roles,
transverse kinetic terms
include coupling between $\Psi_+$ and $\Psi_-$ fields,
and break the charge-color separation.
This is consistent with the fact that in dimensions
larger than one,
increase of quark density affects
the color sector through screening effects.

\section{Summary}
\label{summary}

In this paper, we have utilized QCD$_2$ 
to illustrate some concepts of Quarkyonic matter.
While Quarkyonic matter should differ from conventional quark matter
in the excitation modes,
it should be distinguished from nuclear matter
by bulk quantities.
These aspects can be seen in quark matter in QCD$_2$,
as a consequence that
quark chemical potential acts very differently on
quark number density and color density.

The nonperturbative dynamics near the Fermi surface
play deterministic roles to classify
the phase structure, 
since bulk contributions computed by
weak coupling methods
are approximately 
common for different phases.
Whether excitations are
deconfined quarks and gluons, 
or confined hadrons and glueballs,
is a key issue to understand
dynamical phenomena in cold quark matter.

Another relevant topic discussed in this paper
was the inhomogeneous distributions
of the chiral condensates and quark number densities.
In our opinion,
such inhomogeneous descriptions 
have potential relevance since they might
smoothly interpolate the Fermi surface of
Quarkyonic matter and very dense nuclear matter\footnote{To avoid confusions,
we have to emphasize that
this crystalization 
is very different from that occured in the Skyrme 
crystals \cite{Klebanov:1985qi,Forkel:1989wc}
or its holographic QCD version \cite{Rho:2009ym}.
The latter appears at very low density,
because of long range attractive potential of $O(\Nc)$
much larger than kinetic energy of $P^2/2M_N \sim 1/\Nc$
(Ref. \cite{Hidaka:2010ph} conjectured
that this problem arises from the overestimated
nucleon axial charge $g_A$ in the $\Nc$ counting).
This description contradicts with 
the real nuclear matter at low density,
which are liquid-like rather than solid-like.
In contrast, here we are discussing 
high density where repulsive hard cores 
of nucleons of $O(\Nc)$ start to overlap,
and nucleons can not move easily.
This is the region where the nucleon Fermi sea starts to transform to
the quark Fermi sea.
The understanding of this region remains a difficult problem.
}.
In both phases, excitations are confined,
and chiral symmetry is broken.

Several other important effects have not been addressed.
In particular, we have not discussed
the color superconductivity \cite{Alford:2007xm}
for which a number of colors are important.
Since the formation of colored diquark condensates
competes with the presence of confining forces,
they must be taken into account simultaneously
for more realistic considerations than those given
in this paper.

\section*{Acknowledgments}
The author would like to give special thanks to 
Y. Hidaka, L. McLerran, R.D. Pisarski, and A.M. Tsvelik
during the collaborations related to this work.
He also acknowledges
G. Basar, 
D. Blaschke,
M. Buballa,
A. Cherman,
T. Cohen,
G.V. Dunne,
E.J. Ferrer,
K. Fukushima,
L.Y. Glozman,
K. Hashimoto,
T. Hatsuda,
V. Incera,
T. Izubuchi,
D.B. Kaplan,
H.K. Lee,
S. Nakamura,
J.M. Pawlowski,
M. Rho, 
B.J. Schaefer,
S-J. Shin, 
E. Shuryak,
D.T. Son,
and
I. Zahed
for enlightening discussions
and/or critical comments
which have forced the author to reconsider
many basic concepts of Quarkyonic matter.
This research is supported
under DOE Contract No. DE-AC02-98CH10886
and Posdoctoral Research Program of RIKEN.



\begin{thebibliography}{99}

\bibitem{McLerran:2007qj}
L.~McLerran and R.~D.~Pisarski,
 \npa{796}{83}{2007}.

\bibitem{McLerran:2008ua}
  L.~McLerran, K.~Redlich and C.~Sasaki,
  Nucl.\ Phys.\  A {\bf 824}, 86 (2009)
  [arXiv:0812.3585 [hep-ph]];
%
A.~Andronic {\it et al.},
 \npb{837}{65}{2010};
%
Y.~Hidaka, L.~D.~McLerran and R.~D.~Pisarski,
 \npb{808}{117}{2008};
%
T.~Brauner, K.~Fukushima and Y.~Hidaka,
 \prd{80}{074035}{2009}
  [Erratum-ibid.\  D {\bf 81}, 119904 (2010)];
%
K.~Miura, T.~Z.~Nakano and A.~Ohnishi,
 \ptp{122}{1045}{2009};
%
S.~Hands, S.~Kim and J.~I.~Skullerud,
 \prd{81}{091502}{2010};

\bibitem{Kojo:2009ha}
  T.~Kojo, Y.~Hidaka, L.~McLerran and R.~D.~Pisarski,
  Nucl.\ Phys.\  A {\bf 843}, 37 (2010)
  [arXiv:0912.3800 [hep-ph]].

\bibitem{Kojo:2010fe}
  T.~Kojo, R.~D.~Pisarski and A.~M.~Tsvelik,
  Phys.\ Rev.\  D {\bf 82}, 074015 (2010)
  [arXiv:1007.0248 [hep-ph]].


\bibitem{'tHooft:1974hx}
  G.~'t Hooft,
\npb{75}{461}{1974}.

\bibitem{Callan:1975ps}
  C.~G.~Callan, N.~Coote and D.~J.~Gross,
\prd{13}{1649}{1976};
%
  I.~Bars and M.~B.~Green,
\prd{17}{537}{1978}.

\bibitem{Colemanbook}
S.~Coleman, {\it Aspects of Symmetry}, Cambridge University Press.

\bibitem{Affleck:1985wa}
  I.~Affleck,
\npb{265}{448}{1986}.

\bibitem{Zhitnitsky:1985um}
  A.~R.~Zhitnitsky,
  \plb{165}{405}{1985}
; {\it ibid}. 
  \prd{53}{5821}{1996}

\bibitem{Salcedo:1990rw}
  L.~L.~Salcedo, S.~Levit and J.~W.~Negele,
\npb{361}{585}{1991};
%
V.~Schon and M.~Thies,
\plb{481}{299}{2000};
%
L.~D.~McLerran and A.~Sen,
\prd{32}{2794}{1985};
%
P.~J.~Steinhardt,
\npb{176}{100}{1980}.
\bibitem{Armoni:1998ny}
  A.~Armoni, Y.~Frishman, J.~Sonnenschein and U.~Trittmann,
  Nucl.\ Phys.\  B {\bf 537}, 503 (1999)
  [arXiv:hep-th/9805155].

\bibitem{KTtransition}
 V.L. Berezinski, Sov. Phys. JETP. 32 (1970) 493;
  J.M. Kosterlitz and D.J. Thouless, 
   J. Phys. C 6 (1973) 1181;
%
\bibitem{Witten:1978qu}
  E.~Witten,
 \npb{145}{110}{1978}.
%

\bibitem{Coleman:1973ci}
  S.~R.~Coleman,
  Commun.\ Math.\ Phys.\  {\bf 31}, 259 (1973).


\bibitem{DGR1992}
D.V. Deryagin, D.Y. Grigoriev, and V.A. Rubakov,
\ijma{7}{659}{1992}. 

\bibitem{Son2000}
  E. Shuster and D.T. Son, \npb{573}{434}{2000};
%
  B.~Y.~Park, M.~Rho, A.~Wirzba and I.~Zahed,
  \prd{62}{034015}{2000}.

\bibitem{Nickel:2009ke}
  D.~Nickel,
  \prl{103}{072301}{2009};
{\it ibid}. \prd{80}{074025}{2009};
  S.~Carignano, D.~Nickel and M.~Buballa,
  \prd{82}{054009}{2010}.

\bibitem{Rapp:2000zd}
  R.~Rapp, E.~V.~Shuryak and I.~Zahed,
  \prd{63}{034008}{2001}.

\bibitem{Nakano:2004cd}
  E.~Nakano and T.~Tatsumi,
  \prd{71}{114006}{2005};
 S.~Maedan,
  \ptp{123}{285}{2010};
 M.~Sadzikowski and W.~Broniowski,
  \plb{488}{63}{2000};
 M.~Sadzikowski,
  \plb{642}{238}{2006};
 T.~L.~Partyka and M.~Sadzikowski,
  \jpg{36}{025004}{2009};
 T.~L.~Partyka and M.~Sadzikowski,
  {\it ibid}. arXiv:1011.0921 [hep-ph].
  B.~Bringoltz,
  JHEP {\bf 0703}, 016 (2007);
  I.~E.~Frolov, V.~C.~Zhukovsky and K.~G.~Klimenko,
  \prd{82}{076002}{2010}.


\bibitem{Mandelstam:1975hb}
  S.~Mandelstam,
  Phys.\ Rev.\  D {\bf 11}, 3026 (1975).

\bibitem{Novikov:1982ei}
  S.~P.~Novikov,
  Usp.\ Mat.\ Nauk {\bf 37N5}, 3 (1982);
%
  E.~Witten,
  Commun.\ Math.\ Phys.\  {\bf 92}, 455 (1984).

\bibitem{Conformalbook}
P. Francesco, P. Mathieu, and
D. Senechal,
{\it Conformal Field Theory}, Springer.

\bibitem{Tsvelik}
A.M. Tsvelik,
{\it Quantum Field Theory in Condensed Matter Physics},
Cambridge University Press.

\bibitem{Affleck:1985wb}
  I.~Affleck,
  \npb{265}{409}{1986}.

\bibitem{Frishman:1992mr}
  Y.~Frishman and J.~Sonnenschein,
  \phr{223}{309}{1993}.

\bibitem{Yee:2011yn}
  H.~U.~Yee and I.~Zahed,
  JHEP {\bf 1107} (2011) 033
  [arXiv:1103.6286 [hep-th]].



\bibitem{Schon:2000he}
  V.~Schon and M.~Thies,
\prd{62}{096002}{2000};
\bibitem{Bringoltz:2008iu}
  B.~Bringoltz,
 \prd{79}{105021}{2009};
 \ibid{79}{125006}{2009}.

\bibitem{Poggio:1975af}
  E.~C.~Poggio, H.~R.~Quinn and S.~Weinberg,
  Phys.\ Rev.\  D {\bf 13}, 1958 (1976).

\bibitem{Glozman:2008fk}
  L.~Y.~Glozman,
  Phys.\ Rev.\  D {\bf 79}, 037504 (2009)
  [arXiv:0812.1101 [hep-ph]].

\bibitem{Freedman:1976xs}
  B.~A.~Freedman and L.~D.~McLerran,
  \prd{16}{1130}{1977};
\ibid{16}{1147}{1977};
\ibid{16}{1169}{1977}.
  A.~Kurkela, P.~Romatschke and A.~Vuorinen,
  \prd{81}{105021}{2010}.

\bibitem{Herbst:2010rf}
  T.~K.~Herbst, J.~M.~Pawlowski and B.~J.~Schaefer,
  Phys.\ Lett.\  B {\bf 696}, 58 (2011)
  [arXiv:1008.0081 [hep-ph]].

\bibitem{Fukushima:2010is}
  K.~Fukushima,
  Phys.\ Lett.\  B {\bf 695}, 387 (2011)
  [arXiv:1006.2596 [hep-ph]].

\bibitem{Lottini:2011zp}
  S.~Lottini and G.~Torrieri,
  arXiv:1103.4824 [nucl-th];
G.~Torrieri and I.~Mishustin,
  Phys.\ Rev.\  C {\bf 82}, 055202 (2010)
  [arXiv:1006.2471 [nucl-th]].

\bibitem{Coleman:1974bu}
  S.~R.~Coleman,
  Phys.\ Rev.\  D {\bf 11}, 2088 (1975).

\bibitem{nucl-th/9609012}
  U.~Zuckert, R.~Alkofer, H.~Weigel and H.~Reinhardt,
  Phys.\ Rev.\ C\ {\bf 55} (1997) 2030
  [nucl-th/9609012].
  L.~P.~Gamberg, H.~Weigel, U.~Zuckert and H.~Reinhardt,
  Phys.\ Rev.\ D\ {\bf 54} (1996) 5812
  [hep-ph/9512294].

\bibitem{arXiv:1008.4029}
  J.~-c.~Wang, Q.~Wang and D.~H.~Rischke,
  Phys.\ Lett.\ B\ {\bf 704} (2011) 347
  [arXiv:1008.4029 [nucl-th]].

\bibitem{Adkins:1983ya}
  G.~S.~Adkins, C.~R.~Nappi and E.~Witten,
  Nucl.\ Phys.\  B {\bf 228}, 552 (1983).

\bibitem{Weinbergbook}
S. Weinberg, 
{\it The Quantum Theory of Fields,  vol.2, Cambridge University Press}.

\bibitem{Schnetz:2005ih}
  O.~Schnetz, M.~Thies and K.~Urlichs,
  \anp{321}{2604}{2006};
  G.~Basar and G.~V.~Dunne,
  \prl{100}{200404}{2008};
\ibid{78}{065022}{2008};
%
  \ibid{79}{105012}{2009};
  C.~Boehmer and M.~Thies,
  \prd{80}{125038}{2009};
  \ibid{81}{105027}{2010}.

\bibitem{Vafa:1983tf}
  C.~Vafa and E.~Witten,
  \npb{234}{173}{1984}.

\bibitem{Ebert:2011rg}
  D.~Ebert, N.~V.~Gubina, K.~G.~Klimenko, S.~G.~Kurbanov and V.~C.~Zhukovsky,
  Phys.\ Rev.\  D {\bf 84} (2011) 025004
  [arXiv:1102.4079 [hep-ph]].


\bibitem{Alford:2007xm}
For recent review, 
 M.~G.~Alford, A.~Schmitt, K.~Rajagopal and T.~Schafer,
  Rev.\ Mod.\ Phys.\  {\bf 80}, 1455 (2008)
  [arXiv:0709.4635 [hep-ph]].


\bibitem{Klebanov:1985qi}
  I.~R.~Klebanov,
  Nucl.\ Phys.\  B {\bf 262}, 133 (1985).

\bibitem{Forkel:1989wc}
  H.~Forkel, A.~D.~Jackson, M.~Rho, C.~Weiss, A.~Wirzba and H.~Bang,
  \npb{504}{818}{1989}.

\bibitem{Rho:2009ym}
  M.~Rho, S.~J.~Sin and I.~Zahed,
  \plb{689}{23}{2010};
%
  K.~Y.~Kim, S.~J.~Sin and I.~Zahed,
  \jhep{0809}{001}{2008};
%
  K.~Nawa, H.~Suganuma and T.~Kojo,
  Phys.\ Rev.\  D {\bf 79}, 026005 (2009)
  [arXiv:0810.1005 [hep-th]].

\bibitem{Hidaka:2010ph}
  Y.~Hidaka, T.~Kojo, L.~McLerran and R.~D.~Pisarski,
  Nucl.\ Phys.\  A {\bf 852}, 155 (2011)
  [arXiv:1004.2261 [hep-ph]].

\end{thebibliography}
\end{document}